\documentclass{aa}  

\usepackage{graphicx}
\usepackage{txfonts}
\usepackage{xcolor}
\usepackage[colorlinks=true,
            linkcolor=blue,
            urlcolor=blue,
            citecolor=blue]{hyperref}
\newcommand\rev[1]{{\color{purple}#1}}

\newcommand{\todo}{\ifmmode \text{\color{purple}\Huge{\(\bullet\)}} \else {\color{purple}{\Huge$\bullet$}}\fi}

\newcommand{\eddington}{\lambda_\mathrm{Edd}}

\newcommand{\lsix}{L_{6 \mu \mathrm{m}}}
\newcommand{\lxsoftabs}{L_{0.5-2\mathrm{keV}}^\mathrm{(abs,corr)}}
\newcommand{\lxhardabs}{L_{2-10\mathrm{keV}}^\mathrm{(abs,corr)}}

\newcommand{\lxsoftabsnoncorr}{L_{0.5-2\mathrm{keV}}^\mathrm{(abs,non-corr)}}

\newcommand{\lambdaedd}{\lambda_\mathrm{Edd}}
\newcommand{\NH}{N_\mathrm{H}}
\newcommand{\NHunit}{N_\mathrm{H}/\mathrm{cm}^{-2}}

\begin{document}
\title{The X-ray statistical properties of dust-obscured galaxies detected by eROSITA}
\author{Akatoki Noboriguchi\inst{1}\and
            Kohei Ichikawa\inst{2,3}\and
            Yoshiki Toba\inst{4,5,6}\and
            Tom Dwelly\inst{7}\and
            Kohei Inayoshi\inst{8}\and
            Toshihiro Kawaguchi\inst{9}\and
            Teng Liu\inst{7}\and
            Yuichi Terashima\inst{6,10}\and
            Yoshihiro Ueda\inst{11}\and
            Masayuki Akiyama\inst{12}\and
            Marcella Brusa\inst{13,14}\and
            Johannes Buchner\inst{7}\and
            Kotaro Kohno\inst{15}\and
            Andrea Merloni\inst{7}\and
            Tohru Nagao\inst{6}\and
            Mara Salvato\inst{7}\and
            Hyewon Suh\inst{16}\and
            Tanya Urrutia\inst{17}
            }
\institute{School of General Education, Shinshu University, 3-1-1 Asahi, Matsumoto, Nagano 390-8621, Japan\\
              \email{noboriguchi@astr.tohoku.ac.jp}
              \and
              Global Center for Science and Engineering, Faculty of Science and Engineering, Waseda University, 3-4-1,
Okubo, Shinjuku, Tokyo 169-8555, Japan
              \and
              Department of Physics, School of Advanced Science and Engineering,
Faculty of Science and Engineering, Waseda University, 3-4-1,
Okubo, Shinjuku, Tokyo 169-8555, Japan\\
              \email{kohei.ichikawa@aoni.waseda.jp}  
              \and
              National Astronomical Observatory of Japan, 2-21-1 Osawa, Mitaka, Tokyo 181-8588, Japan
              \and
              Academia Sinica Institute of Astronomy and Astrophysics, 11F of Astronomy-Mathematics Building, AS/NTU, No.1, Section 4, Roosevelt Road, Taipei 10617, Taiwan
              \and
              Research Center for Space and Cosmic Evolution, Ehime University, 2-5 Bunkyo-cho, Matsuyama, Ehime 790-8577, Japan
              \and
              Max Planck Institut für Extraterrestrische Physik, Giessenbachstrasse 1,  Garching bei München 85748, Germany
              \and
              Kavli Institute for Astronomy and Astrophysics, Peking University, Beijing 100871, China
              \and
              Department of Economics, Management, and Information Science, Onomichi City University, Hisayamada 1600-2, Onomichi, Hiroshima 722-8506, Japan
              \and
              Graduate School of Science and Engineering, Ehime University, 2-5 Bunkyo-cho, Matsuyama, Ehime 790-8577, Japan
              \and
              Department of Astronomy, Kyoto University, Kitashirakawa-Oiwake-cho, Kyoto 606-8502, Japan
              \and
              Frontier Research Institute for Interdisciplinary Sciences, Tohoku University, Sendai 980-8578, Japan
              \and
              Dipartimento di Fisica e Astronomia, Università di Bologna, via Gobetti 93/2, Bologna 40129, Italy
              \and
              INAF – Osservatorio di Astrofisica e Scienza dello Spazio di Bologna, via Gobetti 93/3, Bologna 40129, Italy
              \and
              Institute of Astronomy, The University of Tokyo, 2-21-1 Osawa, Mitaka, Tokyo 181-0015, Japan
              \and
              Gemini Observatory/NSF’s NOIRLab, 670 N. A’ohoku Place, Hilo, HI 96720, USA
              \and
              Leibniz-Institut für Astrophysik, Potsdam, An der Sternwarte 16, Potsdam 14482, Germany
              }
\date{Received XXXX XX, XXXX; accepted XXXX XX, XXXX}

\abstract
{The tight relation between a supermassive black hole (SMBH) and host galaxy masses suggests co-evolution of the SMBH with its host galaxy.
Dust-obscured galaxies (DOGs) are considered to be in a co-evolution phase, with the associated active galactic nuclei (AGN) obscured by dust and gas.
Although the DOGs are thought to harbor rapidly growing SMBHs, their X-ray statistical properties, crucial for understanding the properties of obscuring gas as well as the accretion disk state and the hot electron corona  around the SMBHs, remain unexplored due to the combination of the low number density of DOGs and the lack of X-ray surveys achieving both of the wide-area and uniformly high-sensitivity observations.
}
{We construct a sample of X-ray-detected DOGs in the \textit{eROSITA} Final Equatorial Depth Survey (eFEDS) field and examine their X-ray statistical properties.}
{To construct the DOGs sample, we combined data from Subaru/HSC SSP (optical), VIKING (near-infrared), and \textit{WISE} (mid-infrared) all-sky surveys.  We then cross-matched the sample with \textit{eROSITA}-detected sources to select X-ray-detected DOGs.
}
{Our results reveal the discovery of 5738 IR-bright DOGs in the footprint covered by both of the eFEDS and VIKING surveys (60 deg$^2$), with 65 objects identified as X-ray-detected DOGs (eFEDS-DOGs). 
Among them, 41 eFEDS-DOGs show a power-law slope in the near to mid-IR bands (power-law DOGs), indicating dust-obscured AGN. 
The hydrogen column density ($\NH$) suggests that eFEDS-DOGs cover even unobscured AGN, spanning $10^{20} < \NHunit \lesssim 10^{23}$. 
On the other hand, the majority of IR-bright DOGs are not detected by \textit{eROSITA}, suggesting that most IR-bright DOGs are heavily obscured by dust and gas with $\NHunit > 10^{23}$. 
Therefore, eFEDS-DOGs, discovered thanks to the wide-area survey by \textit{eROSITA}, are newly found populations showing less obscured phases among the lifetime of DOGs.
This feature possibly traces the onset of the decrease of the dust and gas obscuration phase of AGN in DOGs; possibly caused by gas stripping/outflows in the AGN feedback phase.
Additionally, some eFEDS-DOGs exhibit deviations, down to nearly 1.0 dex below the monochromatic luminosity at 6 $\mu$m ($\lsix$) versus absorption-corrected intrinsic X-ray luminosity between 0.5--2 keV ($\lxsoftabs$) relation, suggesting that it may signal high Eddington ratios reaching the Eddington limit.
Such a deviation could be realized only when the accretion disk is in a phase of high Eddington ratios, possibly reaching Eddington limit.
This suggests that eFEDS-DOGs are prominent candidates for very rapidly growing black holes reaching close to over the Eddington accretion rate and just beginning the onset of intensive AGN feedback.
}
{}



   \maketitle
%

\section{Introduction}\label{S1}
Since the 1990s, a tight correlation has been observed between the mass of a supermassive black hole (SMBH) and the properties of its host galaxy in the low-redshift universe \citep[e.g.,][]{1998AJ....115.2285M, 2000ApJ...539L...9F, 2000ApJ...539L..13G, 2002ApJ...574..740T, 2003ApJ...589L..21M, 2013ARA&A..51..511K, 2020ApJ...888...37D}. 
This tight correlation implies that SMBHs grow in tandem with the parameters of their host galaxies \citep[e.g.,][]{1998AJ....115.2285M, 2000ApJ...539L...9F, 2000ApJ...539L..13G, 2002ApJ...574..740T, 2003ApJ...589L..21M, 2007ApJ...665..120A, 2013ARA&A..51..511K, 2020ApJ...888...37D}, which suggests co-evolution of the SMBH with its host galaxy.
One scenario proposed to explain the formation and evolution of quasars within the context of co-evolution is the gas-rich major merger scenario \citep[e.g.,][]{1988ApJ...325...74S, 2008ApJS..175..356H, 2012ApJ...758L..39T, 2018PASJ...70S..37G}. 
In this scenario, at least two galaxies undergoing a gas-rich major merger transition into a quasar through a dusty active star-forming (SF) phase and a dusty active galactic nucleus (AGN) phase. 
However, both the SF and AGN phases are expected to be highly obscured by surrounding dust for most of the time, preventing us from observing the most active phases \citep[e.g.,][]{2008ApJS..175..356H}.

By combining Subaru Hyper Suprime-Cam \citep[HSC;][]{2018PASJ...70S...1M}-Subaru Strategic Program \citep[SSP;][]{2018PASJ...70S...4A} wide-field imaging data (optical), VISTA Kilo-degree Infrared Galaxy survey \citep[VIKING;][]{2007Msngr.127...28A} data (near-infrared: NIR), and {\textit{Wide-field Infrared Survey Explorer}} \citep[{\textit{WISE}};][]{2010AJ....140.1868W} all-sky survey \citep[ALLWISE;][]{2014yCat.2328....0C} data (mid-infrared: MIR), \cite{2015PASJ...67...86T, 2017ApJ...835...36T} and \cite{2019ApJ...876..132N} surveyed dust-obscured star-bursting galaxies and/or dusty AGNs \citep[dust-obscured galaxies: DOGs;][]{2008ApJ...677..943D}.
Such DOGs are originally defined as galaxies that are bright in MIR while faint in optical; $(i-[22])_{\rm AB} \geq 7.0$ \citep{2015PASJ...67...86T}. 
Once we adopted the criterion, the number density of DOGs is $\log\phi = -6.59\pm0.11$ [Mpc$^{-3}$] \citep{2015PASJ...67...86T} by assuming $z=1.99\pm0.45$ \citep{2008ApJ...677..943D}.

In the context of the gas-rich major merger scenario, it is expected that the SF-phase evolves into the AGN-phase because the merging event leads to active SF while the gas accretion onto the nucleus caused by a merger requires some time \citep[e.g.,][]{2007ApJ...671.1388D, 2008ApJS..175..356H, 
ich14,2017A&A...608A..90M}. 
Since such active galaxies are heavily obscured by dust and gas, DOGs potentially correspond to actively growing but heavily dust obscured galaxies in the SF-phase or AGN-phase \citep{2008ApJ...677..943D}.
Based on the spectral energy distributions (SEDs) of DOGs between NIR and MIR, DOGs are quantitatively classified into sub-classes, ``bump DOGs'' and ``Power-Law (PL) DOGs'' \citep[see Section~\ref{MIR_CLASS_DOGs};][]{2015PASJ...67...86T, 2019ApJ...876..132N, 2022PASJ...74.1157S, 2022ApJ...936..118Y}. 
The bump DOGs show rest-frame 1.6 $\mu$m stellar bumps, tracing stars at age $>10$~Myr, 
in the SEDs \citep[e.g.,][]{saw02,2009ApJ...700.1190D, 2011ApJ...733...21B}, while the PL DOGs show a power-law feature in the SEDs between optical and MIR, a hallmark feature of AGN \citep[e.g.,][]{2006ApJ...640..167A, 2008ApJ...677..943D, 2008ApJ...672...94F, 2009ApJ...705..184B, 2012AJ....143..125M,lyu22}. 
Therefore, it has been considered that the bump DOGs correspond to the SF-dominant phase in the scenario while the PL DOGs correspond to the AGN-dominant phase.

Among the PL DOGs, some show interesting features that are seemingly contradictory. Although PL DOGs are considered to be dust-reddened or dust-obscured AGN, a certain fraction of PL DOGs shows optical blue excess \citep[blue-excess DOGs: BluDOGs;][]{2019ApJ...876..132N, 2022ApJ...941..195N}, selected based on the power-law index for observed-frame optical bands ($g$, $r$, $i$, $z$, and $y$-bands). 
Even though their overall IR SEDs are dust-obscured AGN, they exhibit the (dust-unobscured) quasar-like optical spectrum and even show broad emission lines, and thus the direct measurement of Eddington ratio ($\lambda_{\rm Edd} = L_{\rm bol}/L_{\rm Edd}$, where $L_{\rm bol}$ and $L_{\rm Edd}$ represent bolometric luminosity and Eddington luminosity, respectively) is also applicable.
\cite{2022ApJ...941..195N} found that the Eddington ratio of BluDOGs exceeds one, and these C~{\sc iv} line profiles show a blue tail. 
These results suggest that BluDOGs are in a super-Eddington phase, and have a nucleus outflow.

\begin{figure}
\centering
\includegraphics[width=8.5cm]{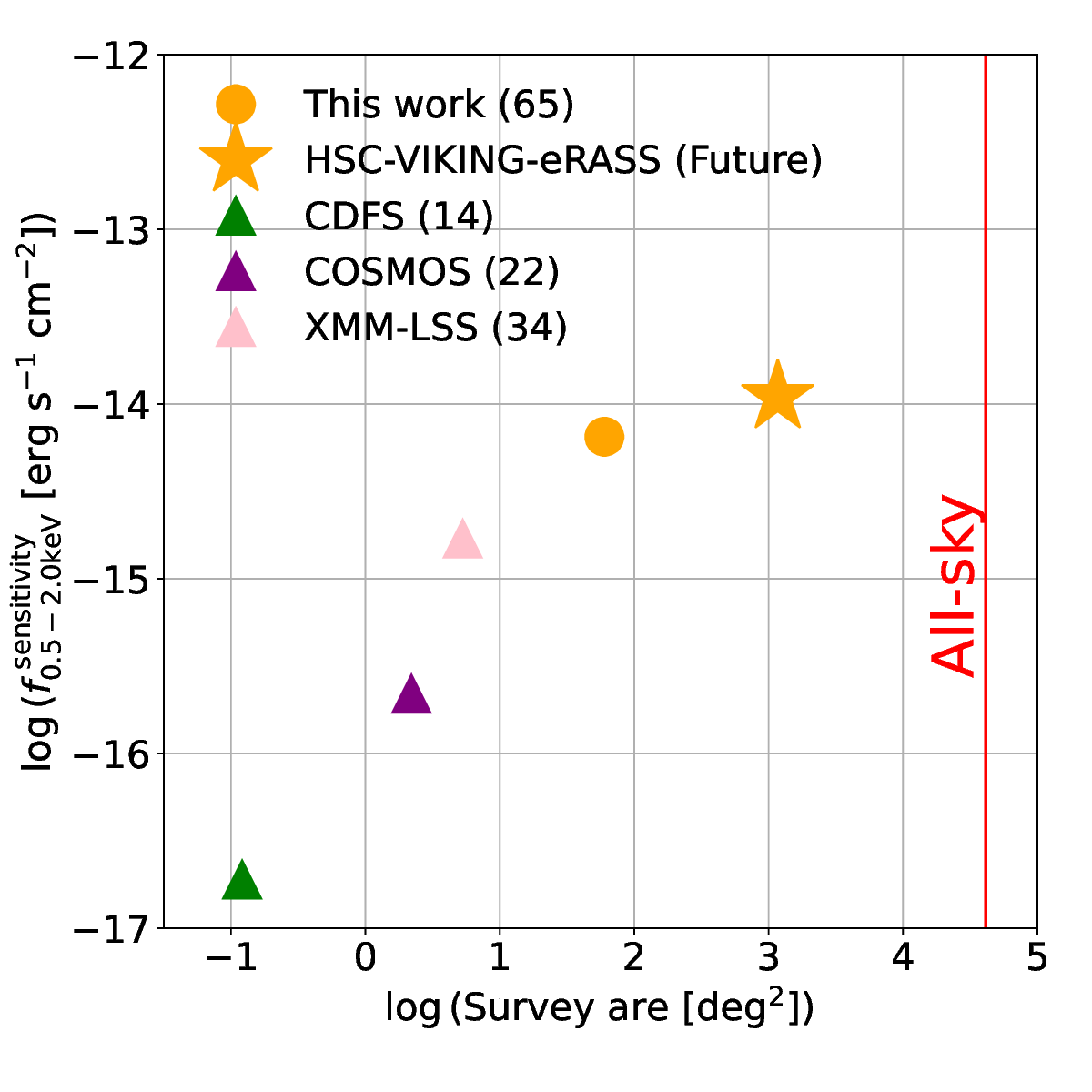}
\caption{Survey area comparison between previous studies and this work. The orange filled circle represents the estimated survey area of this work, while the star plot denotes the final release status of HSC-VIKING-eRASS. The green, purple, and pink triangles denote the survey areas of previous X-ray detected DOG studies in the Chandra Deep Field South (CDFS; \citealt{2016A&A...592A.109C}), the COSMOS field \citep[][]{2019AJ....157..233R}, and XMM-LSS field \citep[][]{2024MNRAS.531..830K}, respectively. The red line outlines the extent of the all-sky area.}
\label{fig:s_area}
\end{figure}

For such obscured objects, X-ray observation data is a powerful tool for researching the physical properties in the obscured region because rest-frame hard X-ray band emission from AGNs are strong against obscurations. 
Given that PL DOGs and BluDOGs are believed to harbor obscured AGNs, X-ray surveys have detected DOGs in the survey footprint, the total number now reaches $\sim100$ sources \citep{2009A&A...498...67L, 2016ApJ...819..111A, 2020ApJ...897..112A, 2016A&A...592A.109C, 2018MNRAS.474.4528V, 2019AJ....157..233R, 2020MNRAS.499.1823Z}.
All DOGs detected by {\textit{Chandra}}, {\textit{XMM-Newton}} \citep{2001A&A...365L...1J}, or {\textit{NuSTAR}} have been found to meet the obscured AGN criterion of hydrogen column density $N_{\rm H} \geq 10^{22}$ cm$^{-2}$, with some classified as Compton thick (CT, $N_{\rm H} \gtrsim 10^{24}$ cm$^{-2}$) AGNs \citep{2016A&A...592A.109C, 2019AJ....157..233R, 2020ApJ...888....8T}.
Recently, \cite{2024MNRAS.531..830K} conducted an X-ray spectral analysis of 34 DOGs using \textit{XMM-Newton} and \textit{Chandra} data in the XMM-LSS field \citep[5.3 deg$^2$: ][]{2018MNRAS.478.2132C,2021ApJS..256...21N, 2024ApJ...977..210Y} as a part of XMM-SERVS survey \citep{pac06,pie16,2018MNRAS.478.2132C}. While a significant fraction ($\sim85\%$) of their X-ray detected DOGs show obscured AGN signatures, they also reported, for the first time, a less obscured AGN population at $10^{21}<\NHunit<10^{22}$.
However, previous X-ray surveys for DOGs were limited to narrow survey areas (< 10 deg$^2$; see Figure~\ref{fig:s_area}), selecting only a fraction of the total DOGs population.

The \textit{extended ROentgen Survey with an Imaging Telescope Array} \citep[\textit{eROSITA};][]{2020NatAs...4..634M,mer24} is the primary instrument on the Spectrum-Roentgen Gamma (SRG) mission \citep[][]{2021A&A...647A...1P, 2021A&A...656A.132S}\footnote{\textit{eROSITA} was launched on July 13, 2019}.
The \textit{eROSITA} Final Equatorial Depth Survey (eFEDS: \citealt{2022AandA...661A...1B}) was released as an X-ray data catalog covering a survey area of $\sim$140 deg$^2$ ($126 < \text{R.A. [deg]} < 146$ and $-3 < \text{Decl. [deg]} < 6$: hereafter eFEDS footprint; \citealt{2022AandA...661A...1B}). 
The eFEDS footprint reaches a flux limit of 
$f_{0.5-2 \mathrm{keV}} = 6.5 \times 10^{-15}$~erg~s$^{-1}$~cm$^{-2}$,
which is 50\% deeper than the final integration of the planned
four-year program (eRASS8) in the ecliptic equatorial region
\citep[$f_{0.5-2 \mathrm{keV}} = 1.1 \times 10^{-14}$~erg~s$^{-1}$~cm$^{-2}$;][]{2021A&A...647A...1P};
 therefore, eFEDS is considered to be a representation of the final eROSITA all-sky survey.

The eFEDS main X-ray catalog contains 27,369 point sources \citep[][]{2022AandA...661A...1B}.
In \cite{2022AandA...661A...3S}, a multi-wavelength catalog for these X-ray sources is constructed (hereafter eFEDS multi-wavelength catalog), and the photometric redshift ($z_{\rm photo}$) of these X-ray sources is estimated using two different methods: machine learning \citep[Nishizawa et al., in prep.][]{} and SED fitting \citep{1999MNRAS.310..540A, 2006A&A...457..841I}. 
For the search of optical and MIR counterparts, DESI Legacy Imaging Surveys Data Release 8 \citep[LS8;][]{2019AJ....157..168D} is adopted. 
LS8 catalog provides not only optical photometry ($g$-, $r$-, and $z$-bands) from the Dark Energy Camera Legacy Survey (DECaLS) but also \textit{WISE} forced photometry from imaging through \textit{NEOWISE-Reactivation} \citep[\textit{NEOWISE-R};][]{2014ApJ...792...30M} measured in the unWISE maps \citep[][]{2014AJ....147..108L, 2016AJ....151...36L} at the locations of the optical sources.
In \cite{2022AandA...661A...5L}, the X-ray spectra are fitted by XSPEC \citep[][]{1996ASPC..101...17A} and BXA \citep[][]{2014A&A...564A.125B, 2021JOSS....6.3045B}, and the X-ray properties (e.g., $\NH$, absorption corrected X-ray intrinsic luminosity between 0.5 keV and 2.0 keV: $\lxsoftabs$) are calculated by adopting the
the spectroscopic redshift ($z_{\rm spec}$) or the estimated $z_\mathrm{phot}$.
Given the medium depth and wide survey area of eFEDS, it provides the first X-ray data for
the statistical sample of DOGs. 
The estimated survey area of the DOG survey with eFEDS is 60 deg$^2$, which is 30 times larger area compared to previous X-ray DOGs survey (see Figure~\ref{fig:s_area}) and the final survey area 
of the combined HSC-SSP optical and the final data release of the \textit{eROSITA} all-sky survey (eRASS) is expected to reach 1000 deg$^2$.

In this paper, the objectives are to select eFEDS-detected DOGs (hereafter eFEDS-DOGs) and report the statistical X-ray properties of eFEDS-DOGs, based on HSC, VIKING, LS8, unWISE, and eFEDS data.
The paper is organized as follows:
We describe the sample selection of DOGs and the classifications in Section~\ref{DA}.
In Section~\ref{R}, we detail the obtained X-ray properties of DOGs, and we discuss the results in Section~\ref{D}.
Conclusions and a summary are provided in Section~\ref{C}.
Throughout this paper, the adopted cosmology is a flat Universe with $H_0 = 70$ km s$^{-1}$ Mpc$^{-1}$, $\Omega_{\mathrm M} = 0.3$, and $\Omega_\Lambda = 0.7$, which are the same as those adopted in \cite{2022AandA...661A...5L} and \cite{2022AandA...661A...3S}. 
Unless otherwise noted, all magnitudes refer to the AB system.

\begin{figure*}
   \centering
   \includegraphics[width=18cm]{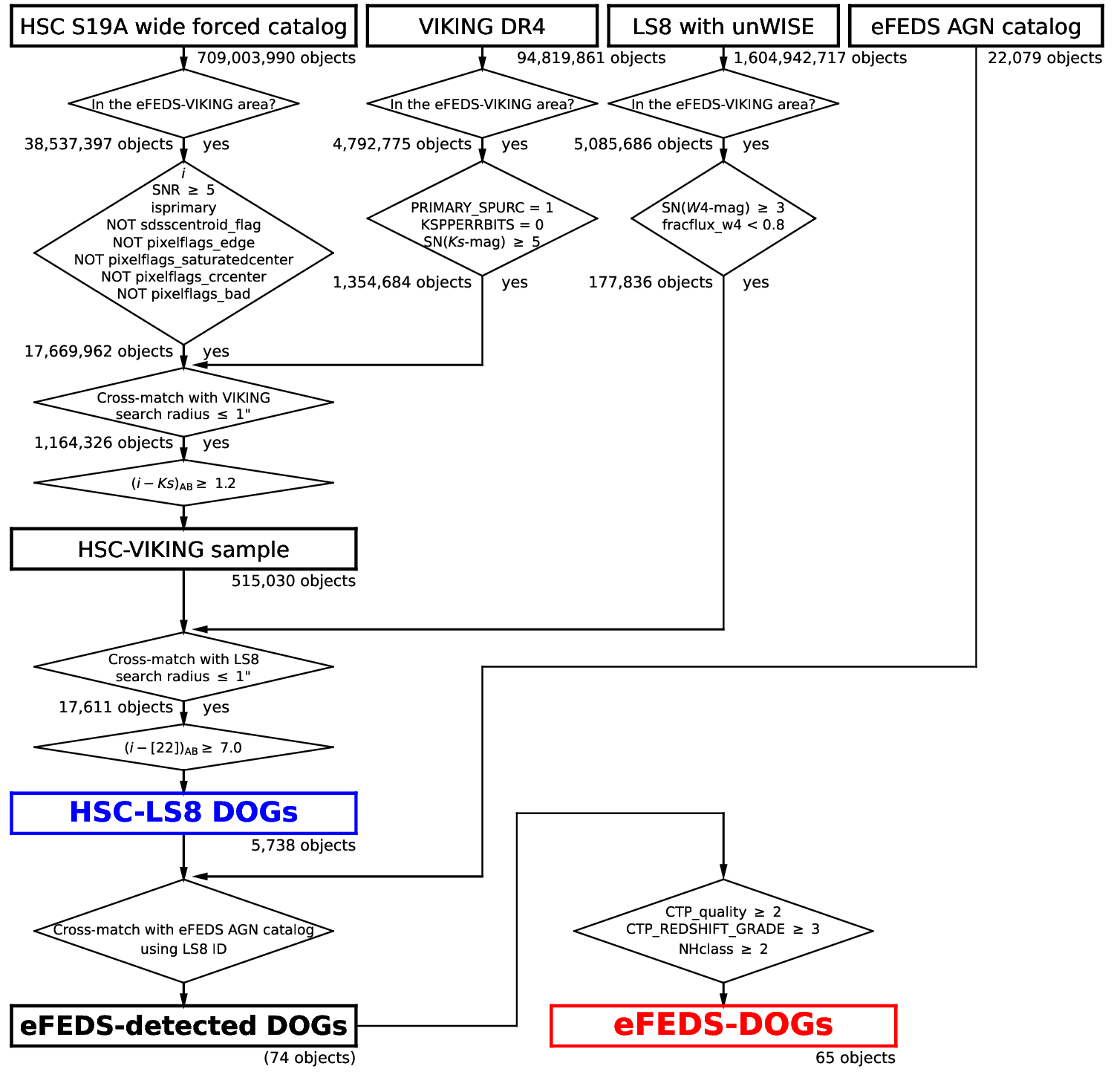}
   \caption{Flow chart of the sample selection for the ``eFEDS-detected DOGs'' (74 objects) and ``eFEDS-DOGs'' (65 objects).}
   \label{fig:fcsc}
\end{figure*}

\section{Data Analysis}\label{DA}

\subsection{Sample selection}\label{DA:SS}

In this study, we selected DOGs detected by {\it eROSITA} in the eFEDS field (eFEDS-DOGs) by combining HSC-DOGs \citep{2015PASJ...67...86T, 2019ApJ...876..132N} with the eFEDS X-ray AGN catalog.
The flow chart of our sample selection process is illustrated in Figure~\ref{fig:fcsc}. 
We identified a total of 65 eFEDS-DOGs and 74 eFEDS-detected DOGs (the definitions will be described later) within an area of approximately 60 deg$^{2}$.

\subsubsection{Catalogs}

In this study, we utilized the HSC catalog (optical), VIKING catalog (NIR), ALLWISE catalog (MIR), Legacy Survey catalog with unWISE photometry data (MIR), and eFEDS AGN catalog (X-ray).

HSC is a wide-field optical imaging camera installed at the prime focus of the Subaru telescope \citep{2018PASJ...70S...3F, 2018PASJ...70...66K, 2018PASJ...70S...2K, 2018PASJ...70S...1M}. 
The S19A catalog, released internally within the HSC survey team, is based on data obtained from March 2014 through April 2019. 
In the eFEDS footprint region, the status of HSC S19A data covers full colors of grizy and full depth, reaching the limiting magnitudes (5$\sigma$, 2\arcsec diameter aperture) with $g$-, $r$-, $i$-, $z$-, and $y$-bands of 26.5, 26.1, 25.9, 25.1, and 24.4 mag, respectively \citep[][]{2018PASJ...70S...8A, 2019PASJ...71..114A}.
In this study, we used a forced photometric catalog of the S19A release \citep[][]{2018PASJ...70S...8A}, analyzed through an HSC pipeline \citep[hscPipe version 7.9.1;][]{2018PASJ...70S...5B} developed by the HSC software team using codes from the Large Synoptic Survey Telescope \citep[LSST;][]{2009arXiv0912.0201L} software \citep[pipline12][]{2008arXiv0805.2366I, 2010SPIE.7740E..15A}.  
Hereafter, we use the cmodel magnitude, estimated by a weighted combination of exponential and de Vaucouleurs fits to the light profile of each object \citep[][]{2001ASPC..238..269L, 2004AJ....128..502A}, to investigate the photometric properties of the sample after correcting for Galactic extinction \citep[][]{1998ApJ...500..525S}.

The VIKING survey is a wide-area NIR imaging survey with five bands ($Z$-, $Y$-, $J$-, $H$-, and $Ks$-bands) observed with the VISTA Infrared Camera on the VISTA telescope \citep[][]{2006SPIE.6269E..0XD}. 
We utilized DR4, the final data release catalog supported by the European Southern Observatory\footnote{http://www.eso.org/rm/api/v1/public/releaseDescriptions/135}. 
The limiting magnitudes (5$\sigma$, 2\arcsec diameter aperture) for the VIKING $Z$-, $Y$-, $J$-, $H$-, and $Ks$-bands are 23.1, 22.3, 22.1, 21.5, and 21.2 mag, respectively. 
We used 2\arcsec-aperture magnitudes in our analysis, and these magnitudes are also corrected for Galactic extinction \citep[][]{1998ApJ...500..525S}.

In this study, we use unWISE data \citep[][]{2019ApJS..240...30S} as MIR data from the LS8 catalog \citep[][]{2019AJ....157..168D}. 
\textit{WISE} observed all-sky in the MIR bands (3.4, 4.6, 12, and 22 $\mu$m). 
The unWISE Catalog, an improvement over the AllWISE catalog \citep[][]{2010AJ....140.1868W, 2013wise.rept....1C, 2014yCat.2328....0C, 2019ApJS..240...30S}, provides deeper imaging and better modeling of crowded regions, detecting sources 0.7 magnitudes deeper than those of AllWISE for bands 3.4~$\mu$m (W1) and 4.6~$\mu$m (W2) \citep[][]{2019ApJS..240...30S}. 
The sensitivities of unWISE catalog at 3.4 ($W1$), 4.6 ($W2$), 12 ($W3$), and 22 ($W4$) $\mu$m-bands are generally better than 19.8, 19.5, 16.4, and 14.5 mag, respectively, although sensitivity depends on the sky position \citep[][]{2010AJ....140.1868W, 2013wise.rept....1C, 2014yCat.2328....0C, 2019ApJS..240...30S}. 
We use profile-fitting magnitude from unWISE photometric information measured at the coordinates of LS8 optical sources.

For X-ray data, we utilized the eFEDS AGN catalog \citep[][]{2022AandA...661A...5L}, which contains the best-estimated redshifts by \cite{2022AandA...661A...3S}. 
\cite{2022AandA...661A...5L} produced the X-ray spectral catalog for 22079 objects with reliable counterparts and with good signal-to-noise.
They extracted the X-ray spectra using \verb|srctool| v1.63 of the eROSITA Science Analysis Software System \citep[eSASS;][]{2022AandA...661A...1B}.
They used multiple models to fit the spectra of the AGN, and the baseline model was an absorbed power-law with \verb|XSPEC| terminology of \verb|TBabs*zTBabs*powerlaw| \citep{2022AandA...661A...5L}.
In addition, for bright objects with a soft excess, they added an additional power-law component to the baseline model, creating a ``double-power-law'' model. 
On the other hand, for faint objects which are too faint to constrain the spectral shape parameters of the ``single-power-law'' model, they fixed the X-ray photon index ($\Gamma$), the absorbing column density ($\NH$), or both at typical values for the sample ($\Gamma=2.0$ and $\NH=0$).
The Galactic absorption (\verb|TBabs|) is also applied using the total $N_\mathrm{H,Gal}$ measured by the neutral HI observations through the HI4PI collaboration \citep{ben16} in the direction of the eFEDS field.
Utilizing the obtained redshift information by \cite{2022AandA...661A...3S}, \cite{2022AandA...661A...5L} derived key AGN properties, including the $\lxsoftabs$, $\NH$, and $\Gamma$.

\subsubsection{Clean Samples for Each Catalogs}\label{SSS_CS}

In selecting DOGs and eFEDS-DOGs, we first establish clean samples for HSC, VIKING, LS8, and eFEDS catalogs. 
For the HSC and VIKING clean sample selection, we refer to the procedures outlined in \cite{2015PASJ...67...86T} and \cite{2019ApJ...876..132N}.

The HSC S19A wide forced catalog initially contains 709,003,990 objects. 
To ensure proper measurements, we exclude objects with inaccurate photometry as follows
(see Figure~\ref{fig:fcsc}). 
Firstly, we select 38,537,397 objects in the overlap region between the eFEDS footprint and VIKING survey area ($129 < \text{R.A. [deg]} < 141$ and $-2 < \text{Decl. [deg]} < 3$).
Consequently, the survey area in this work is limited to the VIKING region within the eFEDS footprint. 
We remove objects affected by the edge, saturated pixels, cosmic rays, or bad pixels using flags such as pixelflags\_edge, pixelflags\_saturatedcenter, pixelflags\_crcenter, and pixelflags\_bad \citep[][]{2018PASJ...70S...5B}.
Objects without a clean centroid measurement in {\textit i}-band are excluded using the flag sdsscentroid\_flag. 
Furthermore, we exclude objects that are not deblended and not unique using the flag isprimary. 
To eliminate objects with unreliable photometry, we exclude those with a signal-to-noise ratio (SNR) < 5 at {\textit i}-band. 
In the end, 17,669,962 objects constitute the HSC clean sample.

The VIKING DR4 catalog initially contains 94,819,861 objects. 
Applying the same region cut as for the HSC clean sample results in 4,792,775 objects. 
We exclude objects that are not unique or are affected by significant noise using flags such as PRIMARY\_SOURCE = 1 and KSPPERRBITS = 0 \citep[][]{2012A&A...548A.119C}. 
Objects with SNR < 5 at {\textit Ks}-band are also removed. 
Consequently, 1,354,684 objects form the VIKING clean sample.

The LS8 catalog\footnote{\url{https://www.legacysurvey.org/dr8/description/}} initially contains 1,604,942,717 objects. 
Applying the region cut as for the HSC clean sample results in 5,085,686 objects. 
Objects whose flux is affected by other sources are excluded using the flag fracflux\_w4 < 0.8. 
Additionally, objects with SNR < 3 at {\textit W4}-band are removed. 
In the end, the LS8 clean sample contains 177,836 objects.

The eFEDS X-ray AGN catalog initially contains 22,079 objects. 
We exclude objects with a counterpart quality that is not good in multi-wavelength analysis \citep{2022AandA...661A...3S} using the flag CTP\_quality $\geq$ 2. 
Objects with spectroscopic or photometric redshifts, estimated by two different methods (SED fitting and machine learning; \citealt{2022AandA...661A...3S}), and with consistent results from both methods are retained using the flag CTP\_REDSHIFT\_GRADE $\geq$ 3. 
Additionally, objects with unreliable $N_{\rm H}$ estimations are excluded using the flag NHclass $\geq$ 2 (see \citealt{2022AandA...661A...5L}). 
In the end, the selection above leaves 17,502 objects form the eFEDS AGN clean sample.
However, to investigate X-ray detection rate, we apply the above flags after cross-matched HSC-LS8 DOGs with eFEDS AGN catalog.

\subsubsection{Cross-matching and Selection}

To minimize mis-identification between HSC and unWISE samples (given their typical angular resolutions of $\sim$0\farcs6 in HSC $i$-band and $\sim$10\arcsec\ in {\it WISE} $W4$-band), we employ an optical-NIR color cut since the VIKING survey has
spatial resolution of $1.0$~arcsec, which is similar one with optical HSC bands  (see \citealt{2015PASJ...67...86T, 2019ApJ...876..132N}).

Firstly, the cross-matched sample between HSC and VIKING contains 1,164,326 objects using the nearest match with a search radius of 1.0 arcsec. 
After that, we apply the color cut ($(i-Ks)_{\rm AB}\geq 1.2$; \citealt{2015PASJ...67...86T, 2019ApJ...876..132N}), and the HSC-VIKING sample comprises 515,030 objects.

For cross-matching the HSC-VIKING sample with the LS8 clean sample with unWISE data, we adopt a search radius of 1.0 arcsec between the coordinates of HSC and LS8 since LS8 coordinates are determined from optical data. 
This process results in the selection of 17,611 objects (HSC-LS8 sample). 
To identify DOGs, we apply the DOGs criterion of $(i-[22])_\mathrm{AB}>7.0$ to the HSC-LS8 sample. 
Consequently, we select 5,738 HSC-LS8 DOGs.

We select eFEDS-detected DOGs from HSC-LS8 DOG sample by LS8 ID cross-matching with the sFEDS AGN catalog.
A total of 74 eFEDS-detected DOGs are selected.
The detected objects with NHclass $<$ 2 are 9 objects, and all of them have CTP\_REDSHIFT\_GRADE $=$ 3. 
Hereafter, we treat those objects as eFEDS-detected DOGs with NHclass $<$ 2, but we do not include them for the remaining discussion using $\lxsoftabs$ and $N_{\mathrm{H}}$.

From the eFEDS-detected DOGs, we further apply additional cut to obtain the sample with reliable X-ray based parameters.
We apply the flags (i.e., CTP\_quality $\geq$ 2, CTP\_REDSHIFT\_GRADE $\geq$ 3, and NHclass $\geq$ 2), and 
a total of 65 objects are selected. Hereafter we call them ``eFEDS-DOGs''. This eFEDS-DOGs sample is the primary one for the results using the X-ray properties, while the eFEDS-detected DOGs sample is used for the results not using the X-ray properties.
This number difference slightly affects the eFEDS detection rate, which will be discussed in Section~\ref{R_UDR}.

Although our sample is based on good redshift quality selection (CTP\_REDSHIFT\_GRADE $\geq$ 3), it is worth noting on the available spectroscopic redshift information. Eight objects out of total 65 sources were observed by SDSS spectroscopic follow-up observations \citep[DR18;][]{alm23} after the construction of eFEDS AGN catalog, but we did not update those information at this stage and we will discuss their spectroscopic properties in the forthcoming paper. Note that six out of 8 objects have correct spec-$z$ confirmation ($|z_{\mathrm{spec}} - z_{\mathrm{photo}}|<0.05$) but 2 objects show $>$ 1$\sigma$ deviation of photo-$z$ uncertainty ($|z_{\mathrm{spec}} - z_{\mathrm{photo}}|>1\sigma$). We will not include these two objects in the following results using luminosities and $\NH$ since those values could be erroneous ones.

\subsubsection{Hard X-ray detected eFEDS-DOGs Sample}
We also checked whether our HSC-LS8 DOGs have hard X-ray counterparts by utilizing the eFEDS hard band catalog, which is based on the high detection likelihood (DET\_LIKE>10) at the 2.3--5~keV band \citep{2022AandA...661A...1B,nan24}. Since the eFEDS hard-band catalog contains the information of the optical counterparts including the corresponding optical LS8 coordinates, we conducted the LS8 ID matching between the HSC-LS8 DOGs and the hard X-ray band catalog. This leaves one source matching. The source is also one of the 65 eFEDS-DOGs sample (the ID is ID\_SRC(main)=608 or 
ID\_SRC(hard)=439). As expected from the detection of the both X-ray catalog, the sample is moderately obscured source with $\log(\NHunit)=21.1$, and it is located one of the lowest redshift source at $z=0.60$. Since the detection is only one source, we will not discuss the hard X-ray properties of this source, but add one flag ``Hard\_Detection'' and the source is flagged ``Y'' (see Table~\ref{table:1} for more details).

\subsection{MIR classification of DOGs}\label{MIR_CLASS_DOGs}
We categorized our eFEDS-DOG samples into three types (unclassified DOGs, bump DOGs, and PL DOGs; \citealt{2019ApJ...876..132N}) based on the SED, using the classification criterion of \cite{2015PASJ...67...86T}. 
We assumed that the SEDs of eFEDS-DOGs are described by a power-law between NIR and MIR, and we fitted the SED between $W2$-, $W3$-, and $W4$-bands using a power-law function. 
With the fitting result, we calculated the expected $Ks$-band flux density described by the extrapolation from the MIR power-law fit ($f^{\rm fit}_{Ks}$).
21 eFEDS-DOGs with fracflux\_w2 > 0.8, fracflux\_w3 > 0.5, and/or SNR($W2$ and/or $W3$) $< 2$ were removed from the classification sample, and these objects were classified as unclassified DOGs. 
We classified the remaining 44 eFEDS-DOGs into bump DOGs and PL DOGs (see Table~\ref{tab:R_C_eFEDS_DOGs}).

Bump DOGs are selected using the following criterion:
\begin{equation}
\frac{f_{Ks}}{f^{\rm fit}_{Ks}} > 3,
\end{equation}
where $f_{Ks}$ is the observed flux density at the $Ks$-band, selecting 3 sources. 
Subsequently, we classified the remaining 41 eFEDS-DOGs as PL DOGs (i.e., ${f_{Ks}}/{f^{\rm fit}_{Ks}} < 3$). 

\subsection{Optical classification of DOGs}\label{OPTICAL_CLASS_DOGs}
Following the BluDOGs criterion outlined in \cite{2019ApJ...876..132N}, we conducted a search for BluDOGs within the eFEDS-DOG sample. 
We estimated the optical spectral slope ($\alpha^{\rm opt}_{grizy}$) by fitting the optical SEDs (HSC $g$-, $r$-, $i$-, $z$-, and $y$-bands). 
None of our sample fulfilled the BluDOGs classification criterion of $\alpha^{\rm opt}_{grizy} < 0.4$ (Table~\ref{tab:R_C_eFEDS_DOGs}).

However, some SEDs of the eFEDS-DOGs exhibited a flattened SED between the HSC $g$-band and $i$-band, and a power-law SED between optical and MIR bands. 
Consequently, we identified BluDOG-like eFEDS-DOGs using the MIR classification and the optical spectral slope ($\alpha^{\rm opt}_{gri}$) obtained by fitting the optical SEDs (HSC $g$-, $r$-, and $i$-bands). 
Two objects were classified as BluDOG-like, adhering to the criteria of MIR\_CLASS = 2 (PL DOGs; see Tables~\ref{tab:R_C_eFEDS_DOGs} \& \ref{table:1}) and $\alpha^{\rm opt}_{gri} < 0.8$.

\subsection{CIGALE SED Fitting}\label{sec:CIGALE}

We also calculated the rest-frame AGN 6~$\mu$m luminosity ($\lsix$), which is used for the discussion
on the accretion disk and hot electron corona properties by comparing to $\lxsoftabs$ (see Section~\ref{D_ER}).
Although the rest-frame 6~$\mu$m emission is generally dominated by AGN dust \citep[e.g.,][]{2017ApJ...837..145C}, eFEDS-DOGs with high SFR might have non-negligible contribution to 6~$\mu$m from the host galaxy component.
To remove such contamination, we conducted the SED fitting of eFEDS-DOGs by using the Code Investigating GAlaxy Emission \citep[CIGALE;][]{2005MNRAS.360.1413B, 2009A&A...507.1793N,2019A&A...622A.103B}.
In this study, we utilized the most up-to-date veresion called the CIGALE 2022.1 \citep{yan22}.
Since the SEDs of eFEDS-DOGs cover the wavelength range from optical $i$-band to WISE W4 ($22~\mu$m), we followed the same model with \cite{2022ApJ...941..195N}, which characterized the SED with the combination of three main components: direct stellar component, host galaxy dust emission (hereafter star formation component), and AGN component (direct AGN emission and AGN heated dust emission).
The details of the parameter setting are summarized in Appendix~\ref{sec:appendix_CIGALE} and in Table \ref{table:CIGALE_parameter_set}.

\begin{figure}
   \centering
   \includegraphics[width=9cm]{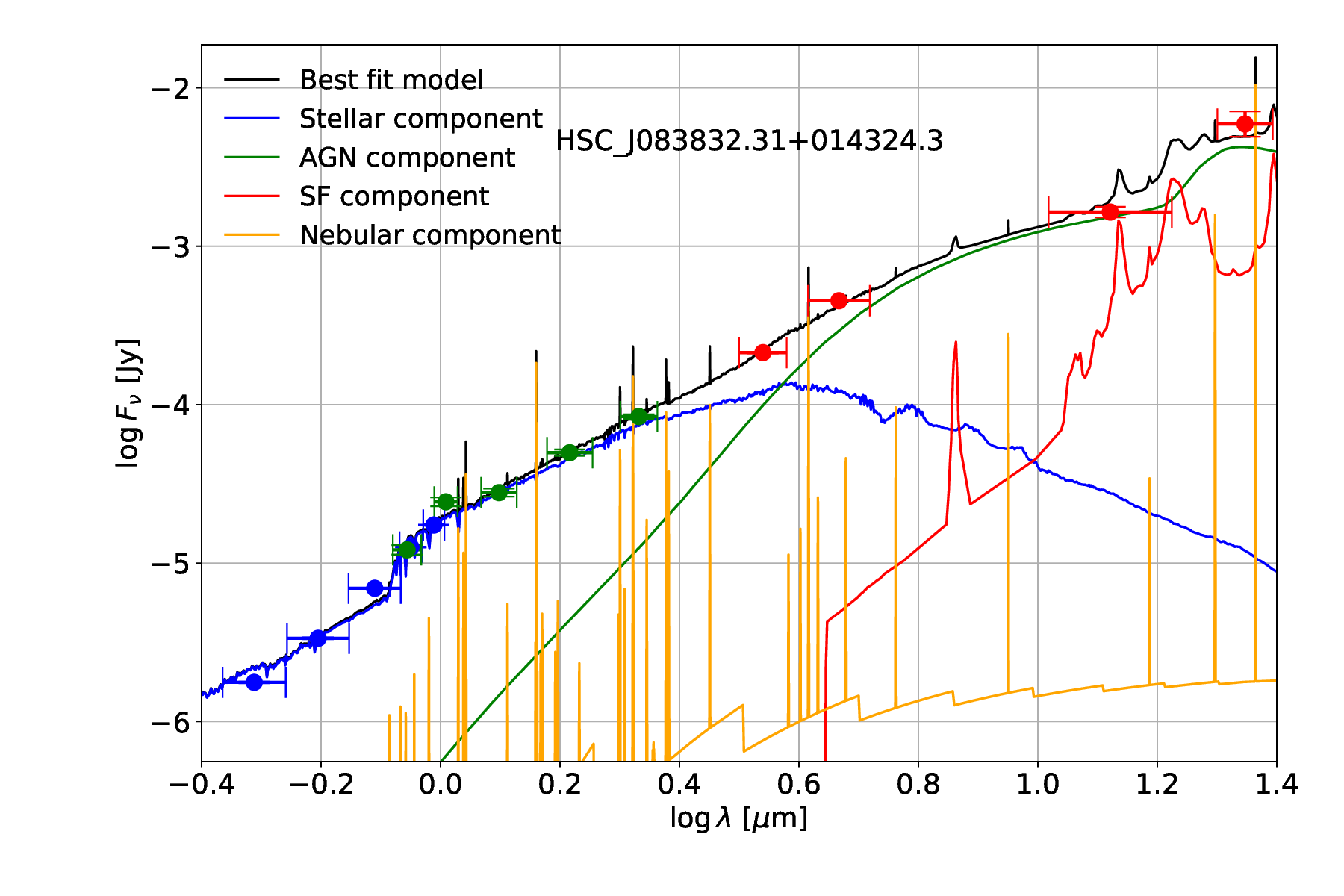}
   \caption{
   Example of the SED fitting result. The blue, green, and red points represent the HSC, VIKING, and WISE photometric data, respectively. The black, blue, green, red, and orange lines correspond to the best-fit model, stellar component, AGN component, star formation (SF) component, and nebular component, respectively
   }
   \label{fig:exampleSED}
\end{figure}

\begin{figure}
   \centering
   \includegraphics[width=9cm]{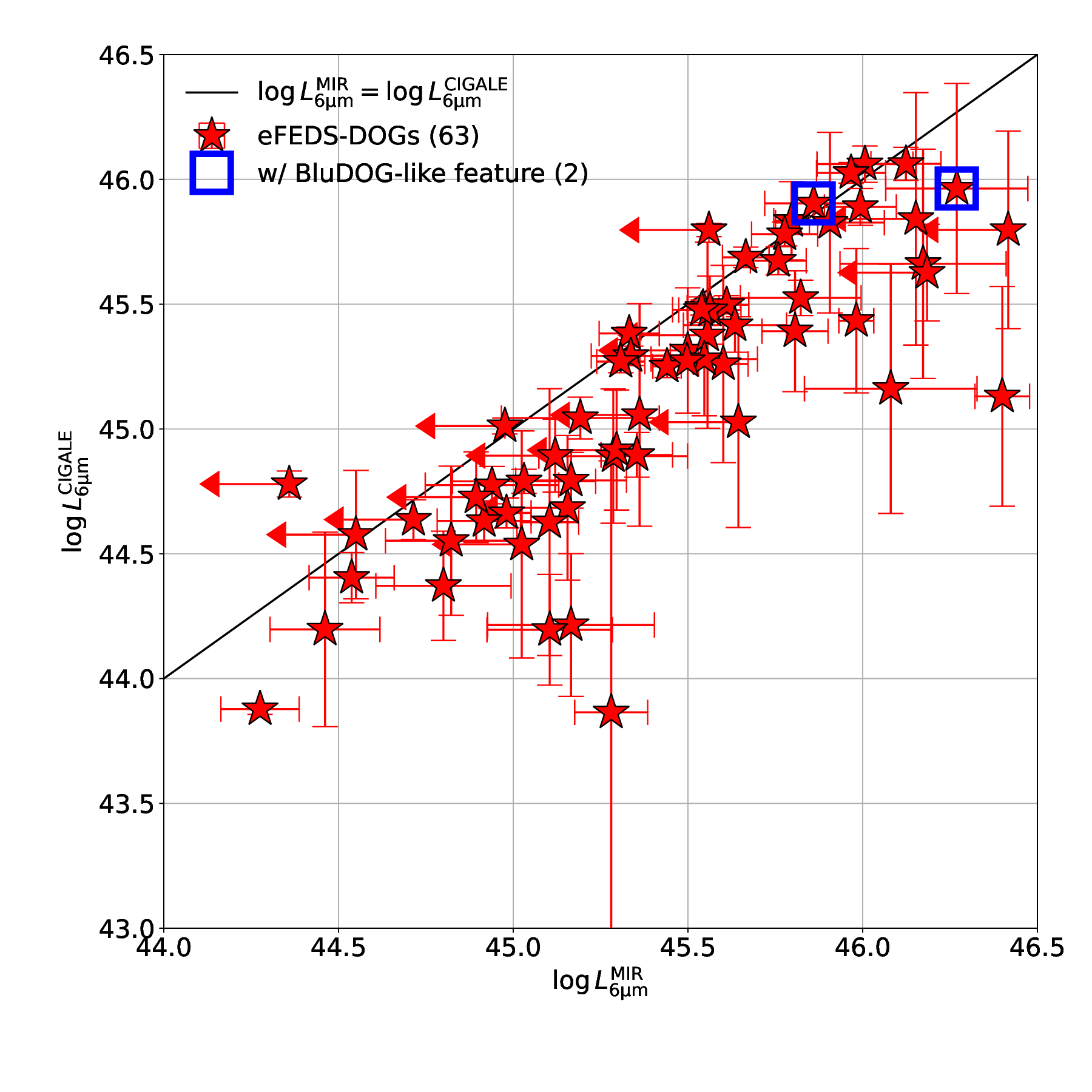}
   \caption{
   $L_{\mathrm{6 \mu m}}$ estimated by CIGALE ($L_{\mathrm{6 \mu m}}^{\mathrm{CIGALE}}$) versus the rest-frame 6~$\mu$m luminosity ($L_{\mathrm{6 \mu m}}$) interpolated from the nearby WISE W1, W2, and W3 bands ($L_{\mathrm{6 \mu m}}^{\mathrm{MIR}}$). Red stars represent the eFEDS-DOGs, while blue squares indicate the eFEDS-DOGs with BluDOG-like features. The black solid line corresponds to $\log L^{\mathrm{MIR}}{\mathrm{6\mu m}} = \log L^{\mathrm{CIGALE}}{\mathrm{6\mu m}}$.
   }
   \label{fig:MIRvsCIGALE}
\end{figure}

Figure~\ref{fig:exampleSED} shows the examples of the SED fitting, and the modeled spectra nicely reproduce the observed SEDs in the range of interests, especiallly around rest-frame 6~$\mu$m.
We then estimated the $L_\mathrm{6 \mu m}$ from the decomposed AGN dust component.
Overall, the estimated $L_\mathrm{6 \mu m}$ has a slightly smaller values with 0.28~dex on average (see Figure~\ref{fig:MIRvsCIGALE}), compared to the rest-frame 6~$\mu$m luminosity interpolated from the nearby WISE W1, W2, and W3 bands.

\begin{table*}
\caption{Result of the classification of
DOGs in the eFEDS footprint}
\label{tab:R_C_eFEDS_DOGs} 
\centering    
\begin{tabular}{l r r r r}
\hline\hline  
                            & eFEDS-DOGs    & eFEDS-detected            & eFEDS-undetected DOGs & HSC-LS8 DOGs\\
Type                        &               & HSC-LS8 DOGs              &                       & \\
                            &               & with NHclass $<$ 2        &                       & \\\hline
MIR SED Classification:     &               &                           &                       & \\
\quad Bump DOGs		        & 3	(4.6\%)     & 0 (0.0\%)                 &  520 (9.2\%)          & 523 (9.1\%)\\
\quad Unclassified DOGs     & 21 (32.3\%)	& 6 (66.7\%)                & 4853 (85.7\%)         & 4880 (85.0\%)\\
\quad PL DOGs			    & 41 (63.1\%)   & 3 (33.3\%)                &  291 (5.1\%)          & 335 (5.8\%)\\\hline
\quad Total				    & 65	        & 9                         & 5664                  & 5738\\\hline\hline
Optical SED Classification: &               &                           &                       & \\
\quad BluDOGs			    & 0	(0.0\%)     & 0 (0.0\%)                 & 9 (0.2\%)             & 9 (0.2\%)\\
\quad BluDOG-like		    & 2	(3.0\%)     & 1 (11.1\%)                & 7 (0.1\%)             & 10 (0.2\%)\\\hline
\end{tabular}
\end{table*}

\begin{figure}
   \centering
   \includegraphics[width=9cm]{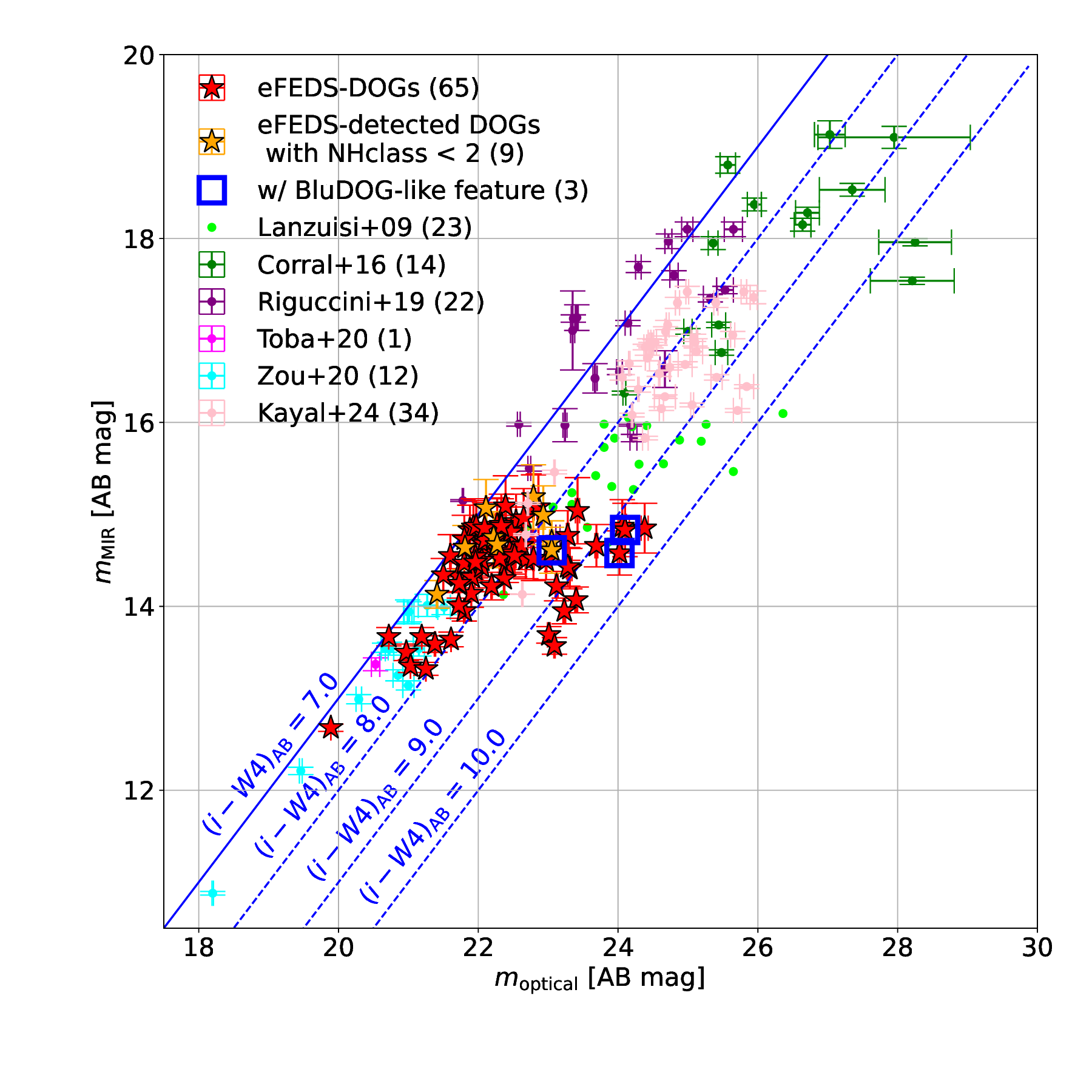}
   \caption{MIR band versus optical band magnitudes of eFEDS-DOGs as well as previous X-ray detected DOGs. Red and orange stars denote eFEDS-DOGs and eFEDS-detected DOGs with NHclass$<$2, respectively. Blue squares show eFEDS-DOGs with the BluDOG-like feature. Lime, Green, purple, magenta, cyan, and pink plots denote X-ray detected DOGs from \cite{2009A&A...498...67L}, \cite{2016A&A...592A.109C}, \cite{2019AJ....157..233R}, \cite{2020ApJ...888....8T}, \cite{2020MNRAS.499.1823Z}, and \rev{\cite{2024MNRAS.531..830K}}, respectively.
   For the sample fainter than $m_\mathrm{MIR}>15$~mag
   \citep[e.g., ][]{2009A&A...498...67L,2016A&A...592A.109C,2019AJ....157..233R}, we utilized Spitzer/MIPS 24~$\mu$m bands as $m_\mathrm{MIR}$. The optical bands were obtained from the Hubble/ACS F775-band \citep{2016A&A...592A.109C}, Suprime-Cam $i_\mathrm{AB}'$ bands \citep{2019AJ....157..233R} and heterogeneous optical magnitudes by Vizier with the nearest matching with $<3$~arcsec \citep{2009A&A...498...67L}.
   The red solid line represents the DOGs criterion, and three red dashed lines denote $(i-W4)_{\rm AB} = 8.0$, $9.0$, and $10.0$. The numbers in parentheses indicate the number of objects.}
   \label{fig:iw4}
\end{figure}

\section{Results}\label{R}
\subsection{Basic Sample Properties of eFEDS-DOGs}\label{R_eDD}
\subsubsection{Distribution in the magnitude planes of MIR and optical bands}

{Figure~\ref{fig:iw4} shows the distribution of 74 eFEDS-detected DOGs on the {\textit{WISE}} $W4$ ($22~\mu$m)-band and HSC $i_\mathrm{AB}$-band diagram. 
Red, orange and blue stars represent eFEDS-DOGs, eFEDS-detected DOGs with NHclass$<$2, and BluDOG-like eFEDS-DOGs, respectively.}
The other X-ray detected DOGs in previous studies are also overlaid in the same plane, and their total number is 106 sources \citep{2009A&A...498...67L, 2016A&A...592A.109C, 2019AJ....157..233R, 2020ApJ...888....8T, 2020MNRAS.499.1823Z, 2024MNRAS.531..830K}.

Figure~\ref{fig:iw4} shows that the number of X-ray-detected DOGs has almost doubled from 106 to 180, and eFEDS-detected DOGs occupy the unique parameter range with ($i_\mathrm{AB}$, $m_\mathrm{MIR,AB}$) = (22--24, 14--16) mag,
which were not explored in the previous X-ray surveys.
This is thanks to the wide-area coverage of the combination of eFEDS and {\textit{WISE}}, resulting in the detections of relatively brighter (originating from the {\textit{WISE}} W4 limiting magnitude of $m_{W4}<15.5$~mag) and relatively rarer DOGs populations that were missed in the previous X-ray survey with $<10$~deg$^2$
\citep{2009A&A...498...67L, 2016A&A...592A.109C, 2019AJ....157..233R, 2020ApJ...888....8T}.
On the other hand, \cite{2020MNRAS.499.1823Z} 
covers the brightest end among the X-ray detected DOGs
with $i_\mathrm{AB}<22$~mag. This is because their sample selection originates from
the combination of the SDSS-{\textit{WISE}} DOGs and heterogeneous but very wide-area XMM-Newton Slew survey area, whose limiting magnitude of $i_\mathrm{AB} \lesssim 22$~mag \citep{2016ApJ...820...46T}.

Additionally, our sample contains wide range of $(i-[22])_{\rm AB}$ color, covering nine eFEDS-DOGs with extremely red color of $(i-[22])_{\rm AB} \geq 9.0$ mag. 
\cite{2016A&A...592A.109C}, \cite{2019AJ....157..233R}, and \cite{2024MNRAS.531..830K} also exhibit very red color objects similar to our sample, and we will discuss those populations in Section \ref{D_NH}.


\begin{figure}
   \centering
   \includegraphics[width=9cm]{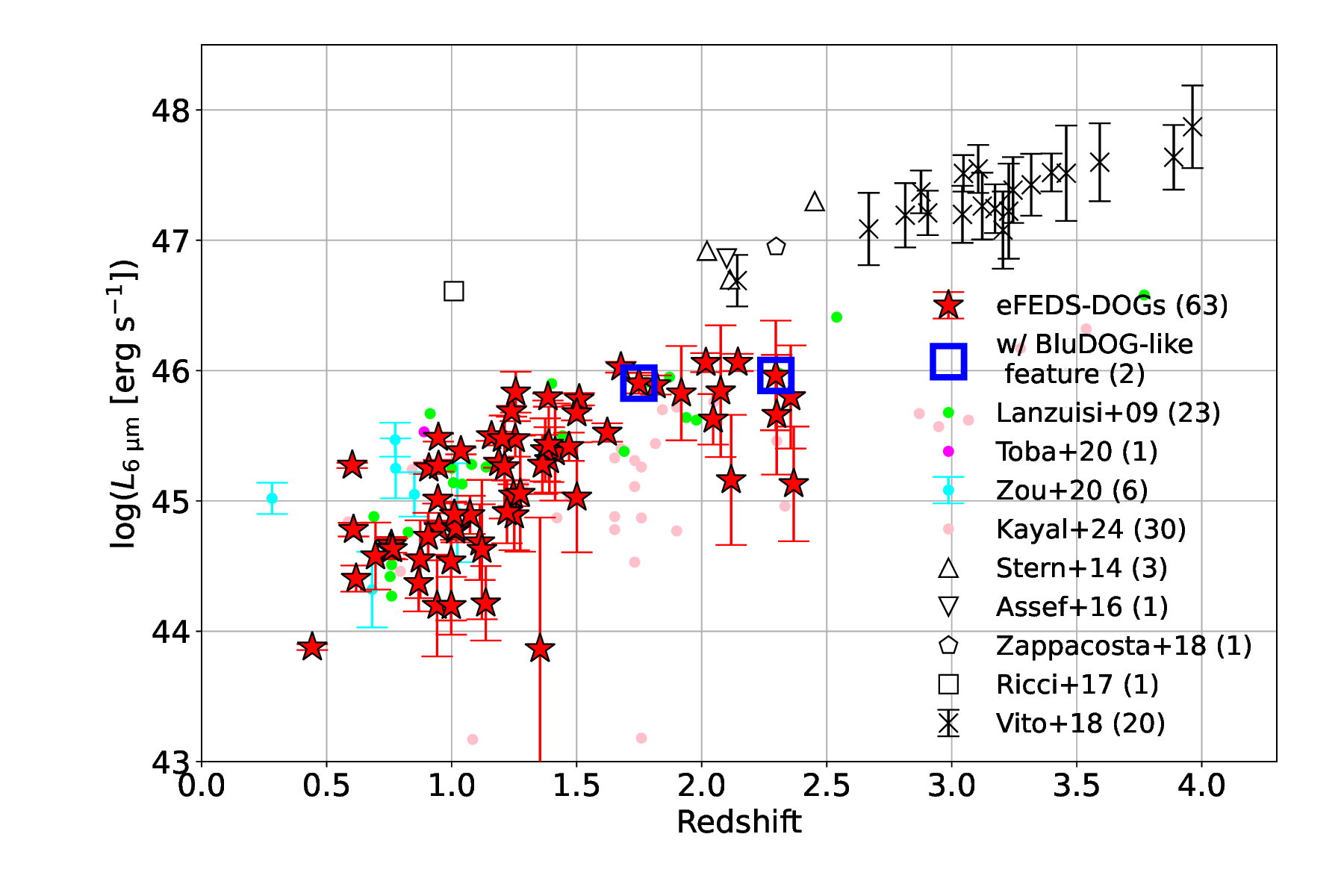}
   \caption{
   The rest-frame 6 $\mu$m AGN luminosity ($L_{\rm 6\ \mu m}$) as a function of redshift. The $L_{\rm 6\ \mu m}$ is estimated from the decomposed AGN dust component as a result of the SED fitting, as discussed in Section~\ref{sec:CIGALE}. The red stars denote the eFEDS-DOGs, and the blue squares denote eFEDS-DOGs with the BluDOG-like feature. The lime, magenta, cyan, and pink plots represent X-ray detected DOGs from \cite{2009A&A...498...67L}, \cite{2020ApJ...888....8T}, \cite{2020MNRAS.499.1823Z}, and \cite{2024MNRAS.531..830K}, respectively. 
   As for the hot DOG samples, black triangle, downward triangle, pentagon, square, and cross represent data from \cite{2014ApJ...794..102S}, \cite{2016ApJ...819..111A}, \cite{2018AandA...618A..28Z}, \cite{2017ApJ...835..105R} and \cite{2018MNRAS.474.4528V}, respectively.}
   \label{fig:L6z_w_os}
\end{figure}

\subsubsection{$L$--$z$ plane}

\begin{table*} 
\caption{Summary of physical parameters}
\label{tab:summaryofpara}      
\centering          
\begin{tabular}{c r c c c c c c}
\hline\hline  
\textbf{Class} & \textbf{Counts} & \textbf{$\log L_{\rm 6\ \mu m}$} & $\log\lxsoftabs$ & $z$ & \textbf{$\log\NH$} & \textbf{$\Gamma$} & $\log\lambda_{\mathrm{Edd}}$ \\
& & $\mathrm{[erg\ s^{-1}]}$ & $\mathrm{[erg\ s^{-1}]}$ & & $\mathrm{[cm^{-2}]}$ & &\\\hline
eFEDS-DOGs          & 63    & $45.2\pm0.6$  & $44.3\pm0.5$  & $1.30\pm0.47$ & $21.7\pm0.8$  & $1.95\pm0.15$ & $-0.63\pm0.26$\\\hline
PL                  & 40    & $45.3\pm0.6$  & $44.3\pm0.5$  & $1.27\pm0.44$ & $21.9\pm0.7$  & $1.94\pm0.15$ & $-0.60\pm0.24$\\
Bump                & 3     & $44.7\pm0.7$  & $44.2\pm0.5$  & $1.48\pm0.58$ & $21.3\pm0.3$  & $1.86\pm0.08$ & $-0.86\pm0.17$\\
Unclassified        & 20    & $45.0\pm0.5$  & $44.2\pm0.4$  & $1.32\pm0.52$ & $21.2\pm0.7$  & $1.98\pm0.14$ & $-0.67\pm0.29$\\\hline
BluDOG-like         & 2     & $46.0\pm0.0$  & $45.0\pm0.1$  & $2.02\pm0.27$ & $22.2\pm0.1$  & $1.85\pm0.01$ & $-0.62\pm0.05$\\
\hline\hline
\end{tabular}
\end{table*}

Figure~\ref{fig:L6z_w_os} shows the rest-frame 6 $\mu$m AGN luminosity
($L_{\rm 6\ \mu m}$) as a function of redshift.
Our eFEDS-DOGs covers wide redshift range covering at $0.5 \lesssim z < 2.5$ (with average of $\left<z\right>=1.30 \pm 0.47$, see Table~\ref{tab:summaryofpara}) and wide luminosity range of $43.7< \log (\lsix/\mathrm{erg}~\mathrm{s}^{-1})<46.2$ (with average of $\left<\log (\lsix/\mathrm{erg}~\mathrm{s}^{-1})\right>=45.2 \pm 0.6$, Table~\ref{tab:summaryofpara}). Figure~\ref{fig:L6z_w_os} also shows the comparison from the sample made by Spitzer Wide-area InfraRed Extragalactic \citep[SWIRE:][]{2003PASP..115..897L} survey \citep[][]{2009A&A...498...67L}, SDSS-WISE survey \citep{2020ApJ...888....8T, 2020MNRAS.499.1823Z}, HSC-MIPS survey \citep{2024MNRAS.531..830K}, and hot DOG surveys \citep{2014ApJ...794..102S, 2016ApJ...819..111A, 2017ApJ...835..105R, 2018AandA...617A..81V, 2018AandA...618A..28Z}. Our sample covers higher-$z$ sources thanks to relatively deeper X-ray follow-ups, as well as the power of deep photometries of Subaru/HSC.

\begin{figure}
   \centering
   \includegraphics[width=9cm]{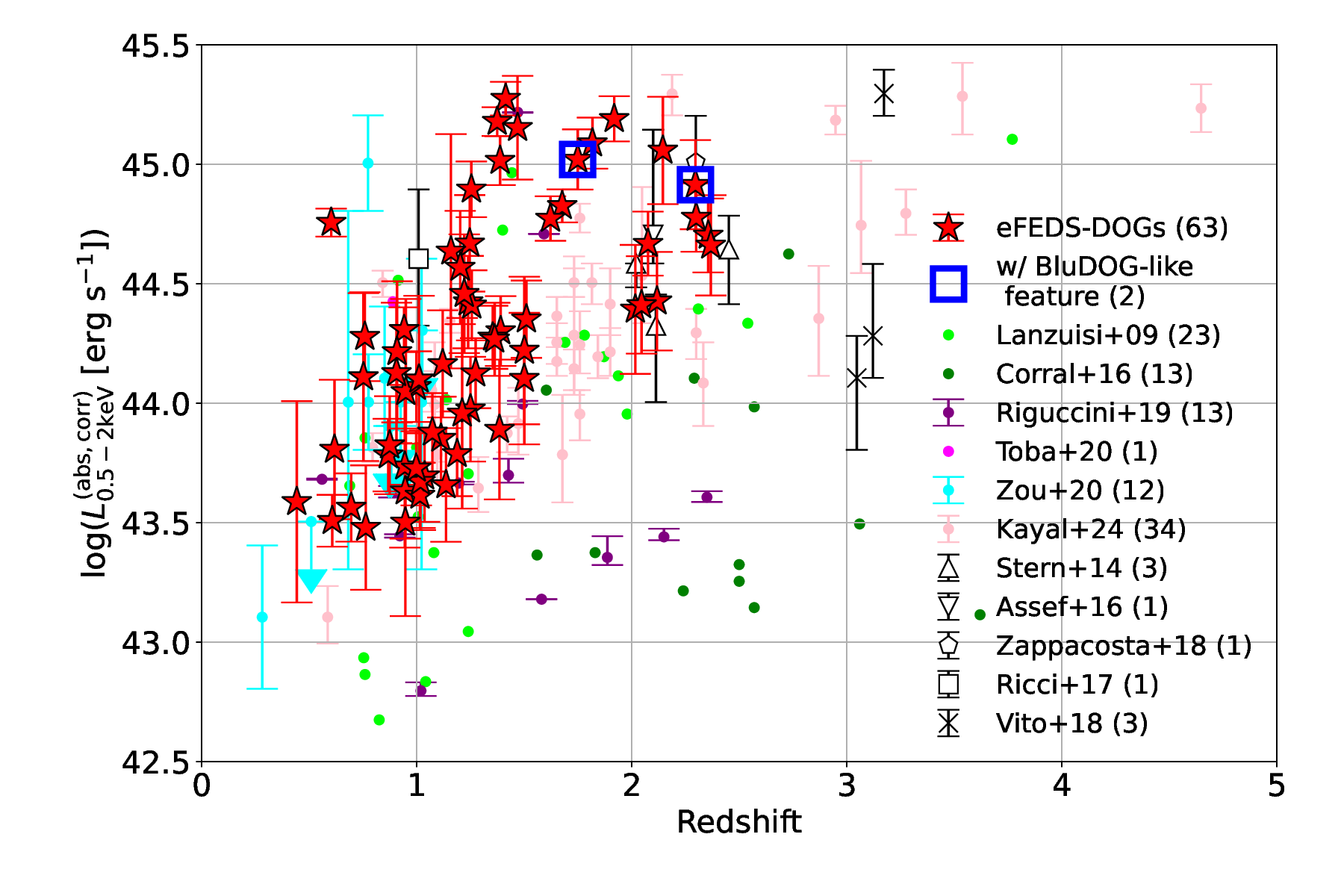}
   \caption{$\lxsoftabs$ as a function of redshift.
   The pionts are same as in Figure~\ref{fig:iw4}. One object presented in \cite{2016A&A...592A.109C} is not shown in this figure due to its low luminosity ($\log (\lxsoftabs/\mathrm{erg}~\mathrm{s}^{-1}) < 42.0$).}
   \label{fig:LXz_w_os}
\end{figure}

Figure~\ref{fig:LXz_w_os} shows the absorption corrected 0.5-2 keV luminosity ($\lxsoftabs$) as a function of redshift. 
For the comparison samples \citep{2009A&A...498...67L, 2016A&A...592A.109C, 2019AJ....157..233R, 2020ApJ...888....8T, 2020MNRAS.499.1823Z, 2024MNRAS.531..830K},
$\lxsoftabs$ were extrapolated from their $\lxhardabs$ values by assuming a photon index of $\Gamma=1.8$ \citep[e.g.,][]{ric17}.
This clearly shows that eFEDS-DOGs cover a relatively luminous end at each redshift. 
Particularly, the eFEDS-DOGs with $z>1$ consists the most luminous X-ray sample among the X-ray detected DOGs, reaching $\lxsoftabs>10^{45}$~erg~s$^{-1}$.

Moreover, the redshift distribution of eFEDS-DOGs covers a wide redshift range
from $z=0.5$ to $z=2.5$, bridging the gap between the low-$z$ sample of $z<1$ \citep{2020ApJ...888....8T,2020MNRAS.499.1823Z} and higher-$z$ sample \citep[$1<z<4$]{2009A&A...498...67L,2016A&A...592A.109C,2019AJ....157..233R}.

Figure~\ref{fig:LXz_w_os} also shows that BluDOG-like eFEDS-DOGs exhibit redshifts between 1.5 and 2.3. This is a natural outcome since the redshifted C~{\sc iv}$\lambda1549$ emission line joins the HSC $g$-bands \citep[][]{2022ApJ...941..195N,2023ApJ...959L..14N}, which makes the optical color bluer within the redshift range of $1.6<z<2.5$.

\begin{figure}
   \centering   
   \includegraphics[width=1.0\linewidth]{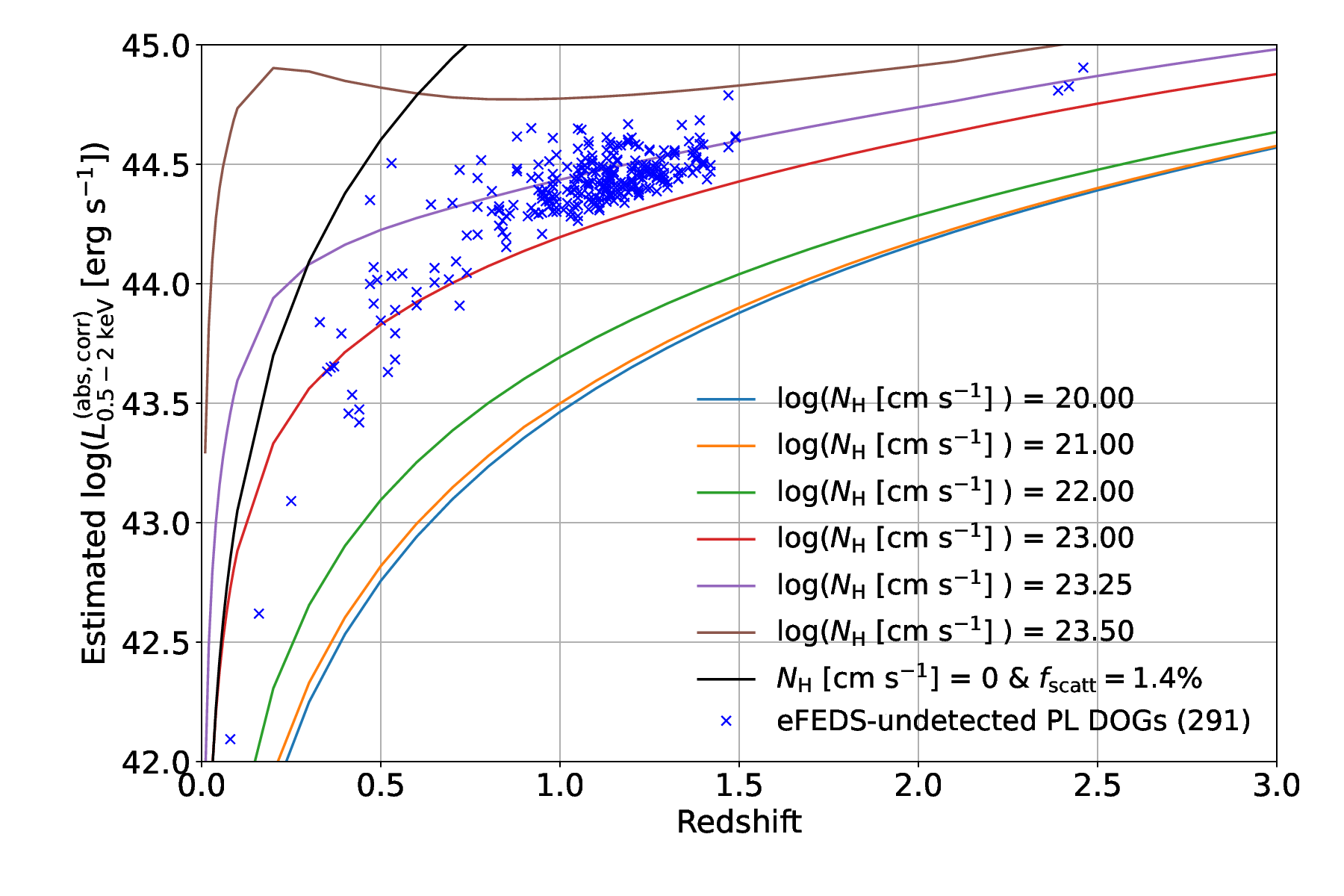}
   \caption{Distribution of the eFEDS-undetected PL DOGs in the plane of $\lxsoftabs$ and redshift.
   The $\lxsoftabs$ were estimated
   from the obtained $\lsix$, by utilizing the luminosity relation between $\lxhardabs$--$\lsix$ \citep[e.g.,][]{2017ApJ...837..145C}, and extrapolated the expected $\lxsoftabs$ by assuming the photon index of $\Gamma=1.8$. 
   The blue crosses represent the eFEDS-undetected PL DOGs. The blue, orange, green, red, purple, and brown solid lines represent the limiting luminosities between 0.5 and 2.0 keV with corresponding 
   $\log (\NHunit)$ of $f20, 21, 22, 23, 23.25,$ and $23.5$, respectively.
   For calculating the limiting luminosities, we utilize a flux limit of $f_{0.5-2{\mathrm{keV}}}=6.5\times10^{-15}$ erg s$^{-1}$ cm$^{-2}$. The black solid line represents the upper bound of the expected X-ray luminosities by assuming that the eFEDS X-ray band is dominated by the X-ray scattered component with $f_\mathrm{scatt}=1.4$\%. This is equivalent to a value that is $100/1.4$ times above the curve of $N_\mathrm{H}$/cm$^{-2}=0$.}
   \label{fig:euDD_Lz}
\end{figure}

\begin{figure}
   \centering
   \includegraphics[width=9cm]{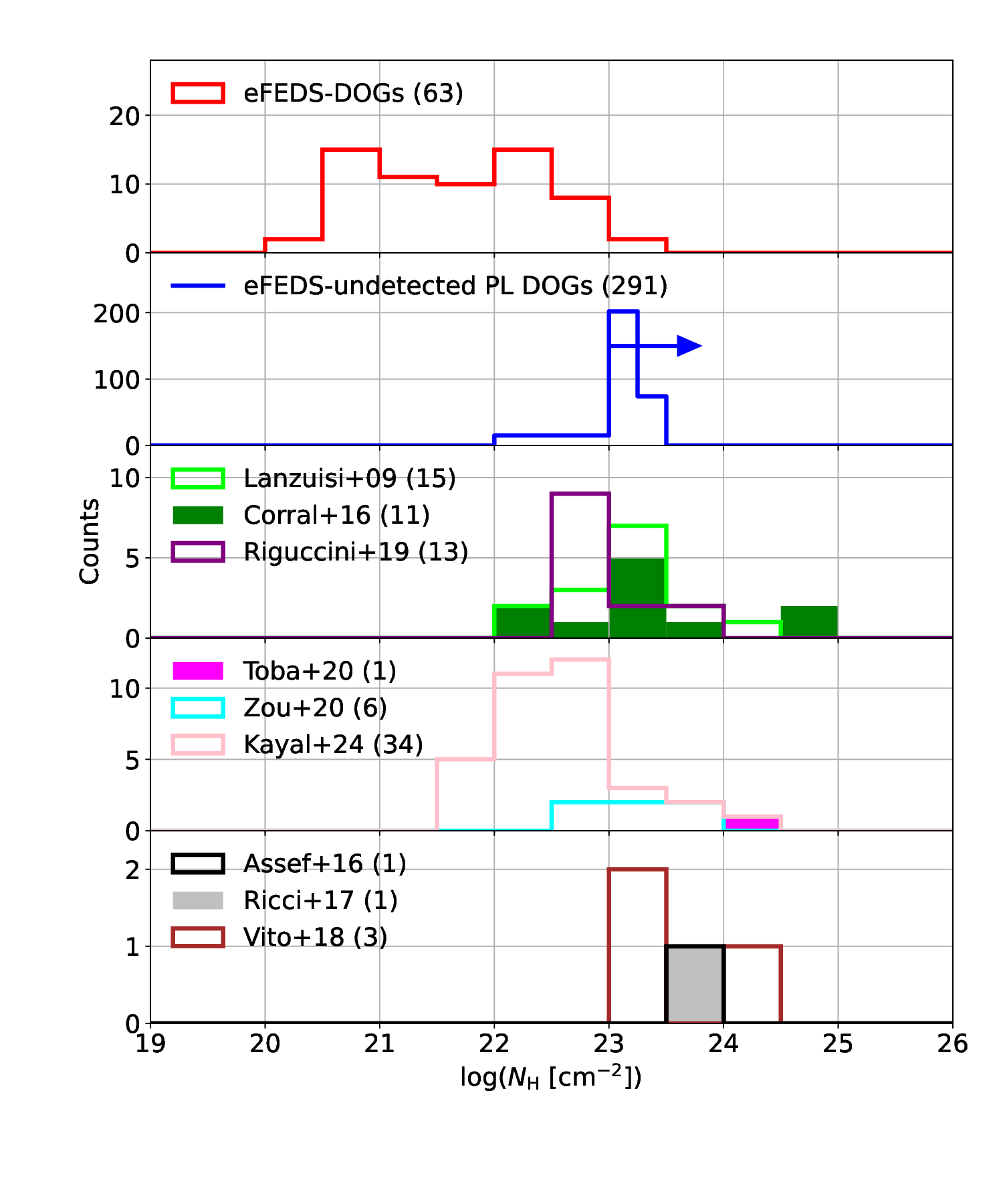}
   \caption{$N_{\rm H}$ histogram with a bin width of 0.5 dex. Top panel: A red histogram represents eFEDS-DOGs. 
   Second panel: A blue histogram represents eFEDS-undetected PL DOGs, indicating lower limits on $\log (\NHunit)$.
   Note that only this panel uses a different binning scheme: 22.0--23.0, 23.00--23.25, and 23.25--23.50.
   Third panel: Lime, green and purple histograms represent \cite{2009A&A...498...67L}, \cite{2016A&A...592A.109C} and \cite{2019AJ....157..233R}, respectively. Fourth panel: Magenta, cyan, and pink histograms represent \cite{2020ApJ...888....8T}, \cite{2020MNRAS.499.1823Z}, and \cite{2024MNRAS.531..830K}, respectively. Bottom panel: black, gray, and brown histograms represent \cite{2016ApJ...819..111A}, \cite{2017ApJ...835..105R}, and \cite{2018AandA...618A..28Z}, respectively. 
   }
   \label{fig:NH_w_os}
\end{figure}

\subsection{Basic Properties of eFEDS-undetected DOGs}\label{R_UDR}

Non-detection by {\textit{eROSITA}} also gives important information either
on the gas obscuration and/or the accretion disk properties of AGN in the DOGs.
The majority of DOGs in the eFEDS footprint is non-detection by {\textit{eROSITA}}, resulting in a low detection rate of 1.3\% (74 sources out of 5738).
Given the energy band of {\textit{eROSITA}} at $E=0.5$--$2.0$~keV,
detection by {\textit{eROSITA}} implies that their column density is $\NH<10^{23}$~cm$^{-2}$ \citep{2007A&A...463...79G},
and the non-detection gives a lower-bound of $\NH$, and most of them are expected to be $\NH > 10^{23}$~cm$^{-2}$, especially at $z>0.5$.

To estimate the lower bound of $N_{\rm H}$ for eFEDS-undetected DOGs,
we utilize the luminosity relation between $\lxhardabs$--$\lsix$ \citep[e.g.,][]{2017ApJ...837..145C}, and extrapolated the expected $\lxsoftabs$ by assuming the photon index of $\Gamma=1.8$. 
The non-detection by {\textit{eROSITA}} also gives the upper-bound of absorption un-corrected 0.5--2~keV luminosity and we estimated the required lower bound of $\NH$ to fulfill the gap between the two values.
In this estimation, we only estimated the $\NH$ values for eFEDS-undetected PL DOGs (291 objects; see Table~\ref{tab:R_C_eFEDS_DOGs}) since PL DOGs have secured AGN features while bump DOGs do not necessarily have AGN in the center. 

Figure~\ref{fig:euDD_Lz} displays the distributions of eFEDS-undetected PL DOGs in terms of estimated $\lxsoftabs$ as a function of redshift. 
These $L_{\rm 6\ \mu m}$ are estimated by a MIR best-fit power-law function between $W2$- and $W4$-bands (see Section~\ref{MIR_CLASS_DOGs}) easily.
However, objects undetected by $W2$- and/or $W3$-bands do not have a best-fit power-law function. 
Therefore, we estimated $L_{\rm 6\ \mu m}$ with/without an upper limit flag, as shown in
Appendix~\ref{sec:appendix_L6} and Table~\ref{tab:L6up}.
The majority of eFEDS-undetected PL DOGs are positioned above the line of $\NH > 10^{23}$~cm$^{-2}$, especially at $z>0.5$. This boundary line is consistent with the expected upper-bound of $\NH$ for {\textit{eROSITA}} detectable sources.

Some might wonder how the non-detection of X-ray ``scattered'' emission as well would give the meaningful constraints to our non-detection. 
The scattered X-ray emission contributes in the soft X-ray bands, especially when the targets are obscured AGN, since the scattered emission originates from the gas-free polar area of the obscuring torus.  
Assuming the average value of the scattered emission of $f_\mathrm{scatt}=1.4$\% \citep{ric17}, the X-ray non-detection gives the upper-bound of the expected X-ray luminosities by a factor of $\approx70$ above the curve of $\NHunit=0$. 
As shown in Figure~\ref{fig:euDD_Lz}, all the esimated $\lxsoftabs$ values are located below the expected upper-bound curve, giving a consistent view.

Figure~\ref{fig:NH_w_os} shows the $N_{\rm H}$ distributions of eFEDS-DOGs, eFEDS-undetected DOGs, and comparison samples.
This demonstrates that the majority of DOGs host heavily obscured AGN with $\NH > 10^{23}$~cm$^{-2}$ by assuming that all DOGs should follow the $\lxhardabs$--$\lsix$ \citep[e.g.,][]{2017ApJ...837..145C}, while eFEDS-DOGs are extremely rare populations among DOGs whose $\NH$ has relatively smaller values with $\NH<10^{23}$~cm$^{-2}$.
Figure~\ref{fig:NH_w_os} also demonstrates that eFEDS-DOGs in this study expand the $\NH$ range for X-ray detected DOGs and eFEDS-DOGs contain even unobscured AGN (gas+dust unobscured DOGs) with $\NH<10^{22}$~cm$^{-2}$, while previous X-ray detected DOGs exhibit $\NH>10^{22}$~cm$^{-2}$, only the obscured AGN 
 \citep{2009A&A...498...67L, 2016A&A...592A.109C, 2019AJ....157..233R, 2020ApJ...888....8T, 2020MNRAS.499.1823Z}.
Among the sample from \cite{2024MNRAS.531..830K}, five objects have $21.5 < \log (N_{\mathrm{H}}/\mathrm{cm}^{-2}) < 22.0$, while the majority show $\log (N_{\mathrm{H}}/\mathrm{cm}^{-2}) > 22.0$. In contrast, the presence of objects with $\log (N_{\mathrm{H}}/\mathrm{cm}^{-2}) < 21.5$ in our sample represents a truly unique population.

Another important possibility for the origin of the X-ray non-detection in eFEDS is that the DOGs are intrinsically X-ray weak. Our results are based on the assumption that AGN in DOGs would follow the $\lxhardabs$--$\lsix$ relation, but some AGN populations lie well below this relation curve. \cite{yam21,yam23} demonstrated that local Ultra-Luminous Infrared Galaxies (ULIRGs) at $z<1$ exhibit extremely X-ray weak features that do not follow the standard $\lxhardabs$--$\lsix$ relation. Similarly, extremely dust-obscured AGN, known as hot DOGs, also exhibit a similar trend at $1<z<3$ \citep{2012ApJ...755..173E, ric17,2018MNRAS.474.4528V,2021A&A...654A..37D}. Given that all these populations (ULIRGs, hot DOGs, and DOGs) are known to display major merger features (\citealt{vei02,ish04,kar10} for ULIRGs and \citealt{2009AJ....137.4854M,2009ApJ...693..750B,2011ApJ...733...21B} for DOGs), our eFEDS non-detected DOGs at $0.5<z<1.5$ might also belong to such intrinsically X-ray weak sources.

If eFEDS-undetected DOGs follow a similar trend as local ULIRGs or hot DOGs at $1<z<3$, the expected $\lxsoftabs$ would decrease by a factor of up to 3 \citep[see][]{ric17,yam23}, and thus their expected lower bounds of $\NH$ would also decrease, with $\NH>10^{22}$ cm$^{-2}$. A deeper hard X-ray observation at $E>2$~keV is crucial to disentangle the degeneracy between obscuration (high $\NH$) and the intrinsic X-ray weak scenario.

In summary, the {\textit{eROSITA}} non-detection provides crucial information suggesting that most eFEDS non-detected DOGs either host heavily obscured AGN with $\NH>10^{23}$~cm$^{-2}$ \citep[see also][]{2022AandA...661A..15T}, exhibit extremely weak X-ray features, or both. On the other hand, a large fraction ($>50$\%) of eFEDS-DOGs show unobscured AGN properties.
{\textit{eROSITA}} opens up such a new population of DOGs, the gas+dust unobscured DOGs that might be in a phase of transition from the heavily obscured to decreased gas-covering fraction, a possible key population tracing the final phase of major merger scenarios {\citep[e.g.,][]{2008ApJS..175..356H}} and/or gas-obscured SMBH growth \citep{ric22,ric23}.

\begin{figure}
   \centering
      \includegraphics[width=9cm]{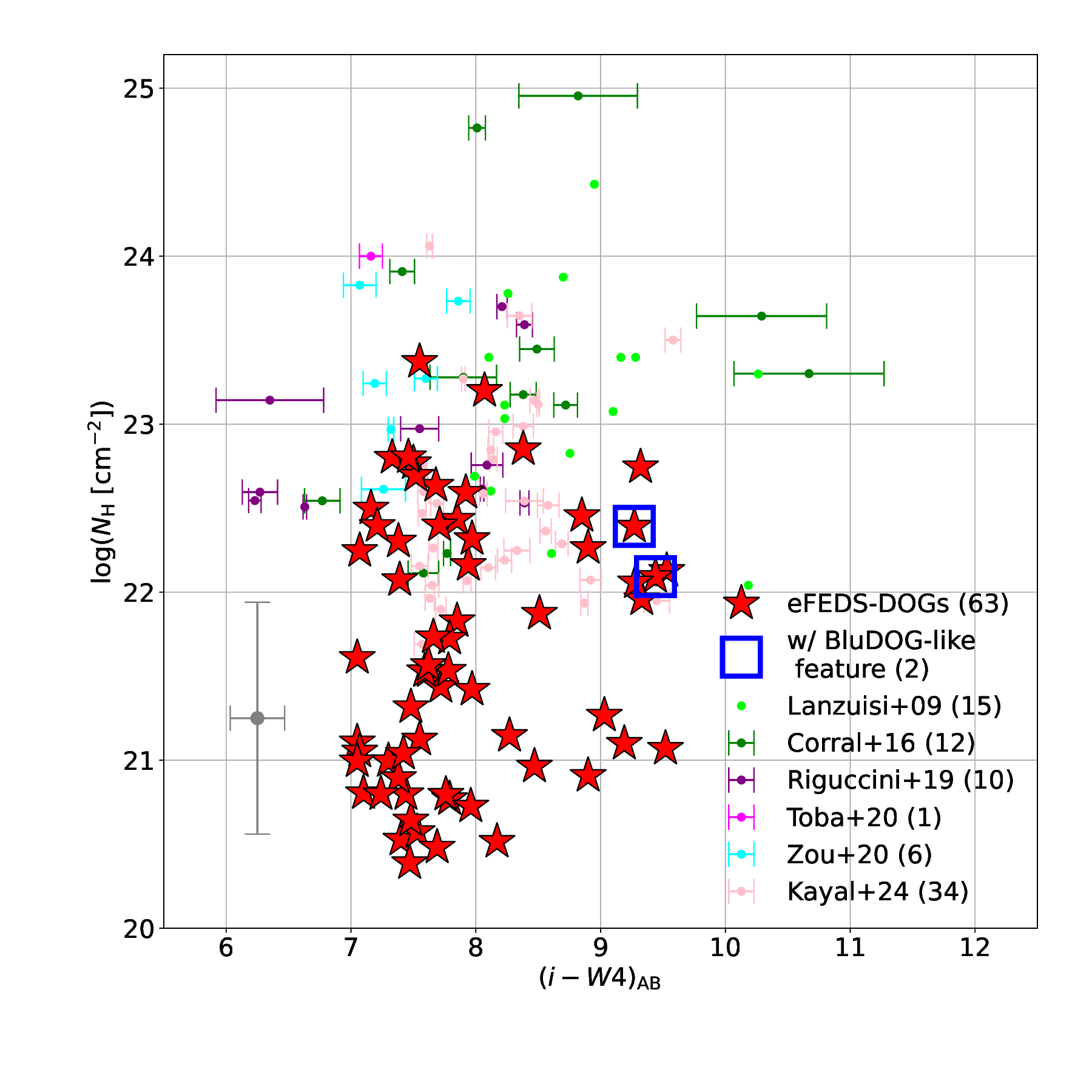}
   \caption{$(i-W4)_{\rm AB}$ vs. $N_{\rm H}$. See Figure~\ref{fig:iw4} for the description of each plots. A gray plot with error bars represents typical standard deviations of $(i-W4)_{\rm AB}$ and $N_{\rm H}$, and the standard deviations of $(i-W4)_{\rm AB}$ and $N_{\rm H}$ are 0.21 mag and 0.69 dex, respectively.}
   \label{fig:iw4nh}
\end{figure}

\section{Discussions}\label{D}

\begin{figure*}
   \centering
   \includegraphics[width=18cm]{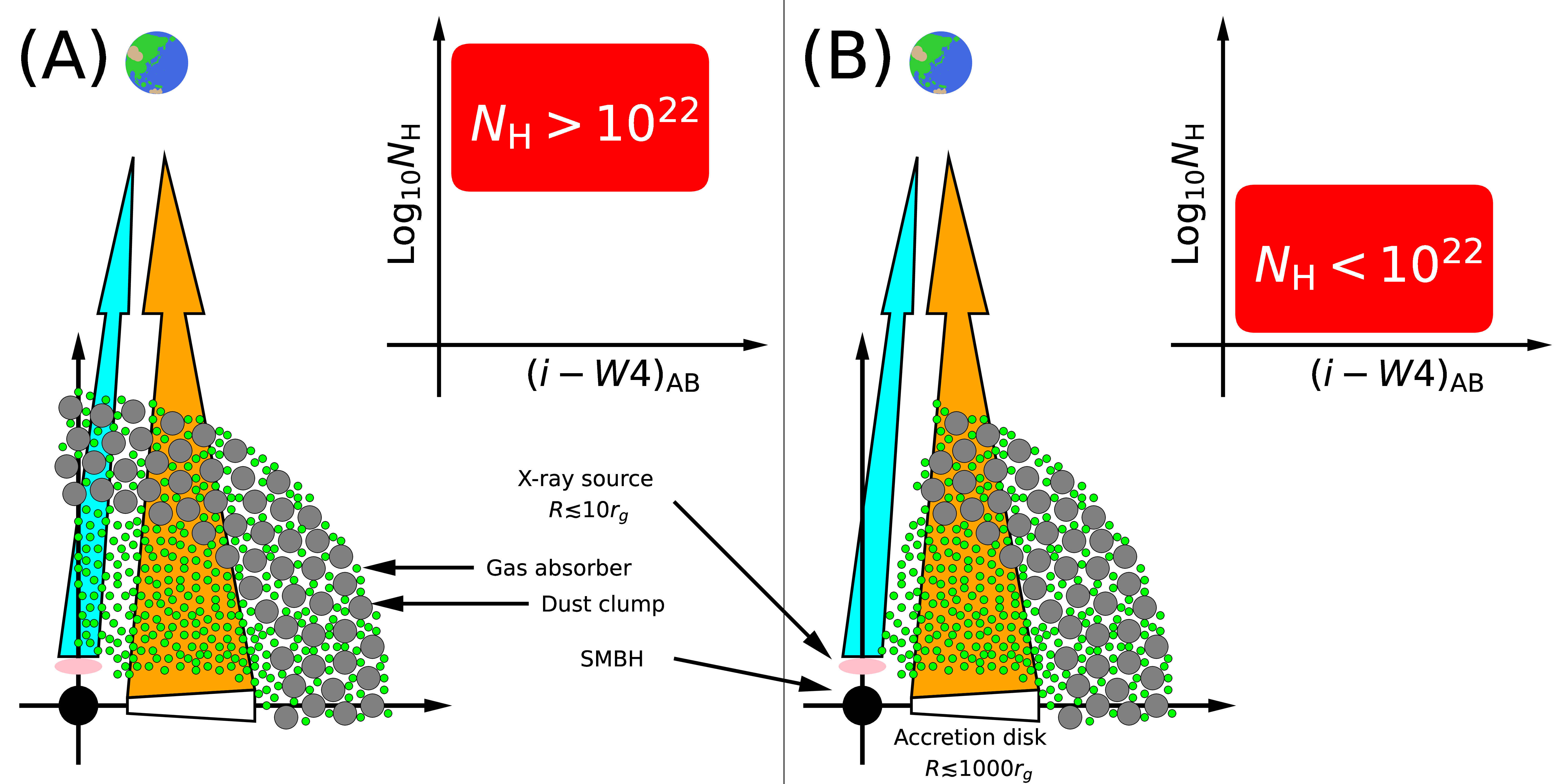}
   \caption{Illustration explaining the $(i-W4)_{\rm AB}$ vs. $N_{\rm H}$ relationship based on geometry. Black, gray, and cyan filled circles denote SMBHs, dust clumps, and gas absorbers, respectively. Pink filled ellipses and white trapezoids represent X-ray sources and accretion disks, respectively. Cyan and orange arrows indicate fluxes from X-ray sources and accretion disks, respectively.}
   \label{fig:image_iw4nh}
\end{figure*}

\begin{table*} 
\caption{The difference between eFEDS-DOGs and eFEDS-undetected DOGs: Physical Parameters, Geometry, and Population}
\label{tab:iw4nh_and_para}      
\centering          
\begin{tabular}{c l l  r r }
\hline\hline  
\textbf{Physical parameter} & \textbf{Explanation} & \textbf{Expression} & \multicolumn{2}{c}{\textbf{Geometry}} \\
 &             & \textbf{in Figure~\ref{fig:image_iw4nh}}   & (A)   & (B)\\
\hline
$N_{\rm H}$ & Gas absorption    & \textcolor[rgb]{0,1,1}{Cyan arrow}            & High & Low\\
$A(i)$      & Dust absorption   & \textcolor[rgb]{1,0.6,0}{Orange arrow}          & High & High\\
$W4$ flux   & Dust re-emission  & \textcolor[rgb]{0.4,0.4,0.4}{Dust covering factor}  & High & High\\
$(i-W4)_{\rm AB}$ & Color       &                       & High                  & High\\\hline
            & Population        &                       & eFEDS-undetected DOGs & eFEDS-DOGs\\\hline\hline
\end{tabular}
\tablefoot{The table summarizes the classification of eFEDS-DOGs based on key physical parameters and their corresponding expressions in Figure~\ref{fig:image_iw4nh}. The ``Geometry'' column indicates the prevalence of high or low values for gas absorption, dust absorption, dust re-emission, and color. The ``Population'' row distinguishes between eFEDS-undetected DOGs and eFEDS-DOGs. The text and color in the ``Expression'' column represent the following: a cyan arrow indicating gas-absorbed flux, an orange arrow indicating dust-absorbed flux, and a gray-filled circle distribution representing the dust covering factor.}
\end{table*}

\subsection{Variations of Gas-to-Dust Ratios}\label{D_NH}

The finding of gas+dust unobscured DOGs with $\NH<10^{22}$~cm$^{-2}$, as shown in Figure~\ref{fig:NH_w_os}, provides several key insights on the gas and dust properties and their geometries for eFEDS-DOGs.
Figure~\ref{fig:iw4nh} illustrates the 
distribution of X-ray detected DOGs in the plane of $\NH$ (gas absorption) and $(i-W4)_{\rm AB}$ (dust extinction).
While previous X-ray detected DOGs are clustered at only obscured AGN region with $\NH>10^{22}$~cm$^{-2}$, the eFEDS-DOGs represent relatively gas+dust unobscured AGN but exhibit an extremely red $(i-W4)_\mathrm{AB}$ color with $7\ \mathrm{[mag]}<(i-W4)_\mathrm{AB}<10\ \mathrm{[mag]}$.

These eFEDS-DOG sources do not demonstrate a clear relationship between $(i-W4)_\mathrm{AB}$ and $\NH$.
One possibility is that the physical locations of media responsible for dust and gas absorption for eFEDS-DOGs might be different.
Such possibilities were reported in several observational studies for AGN obscuration, suggesting the presence of dust-free gas that is responsible for part of the X-ray absorption \citep[e.g., long-term X-ray variabilities;][]{mai10,ris07,ris11,mar14}, 
\citep[the expected location of Fe~K$\alpha$ line emitting material; ][]{min15,gan15}, \citep[e.g., the discrepancy between the gas covering fraction and the dust one; ][]{dav15,ich19}.

On the other hand, our result shows an opposite trend; a lack of gas obscuration into the line-of-sight, but the existence of dust absorption (extinction) on a larger physical scale. 
Figure~\ref{fig:image_iw4nh} illustrates one possible schematic sketch to explain the obtained $(i-W4)_{\rm AB}$ vs. $N_{\rm H}$ relationship for eFEDS-DOGs.
In the left upper panel (A in Figure~\ref{fig:image_iw4nh} and Table~\ref{tab:iw4nh_and_para}), heavy obscuration is caused by both gas and dust. 
These objects exhibit $\log (\NHunit)>23$ and higher $(i-W4)_{\rm AB}$ values, suggesting that both the UV emitting region in the accretion disk and the X-ray emission region are heavily obscured by gas and dust. 
The majority of eFEDS-undetected DOGs, with an estimated lower limit of $\log (\NHunit)>23$, likely exhibit a similar gas and dust geometry as depicted in (A) in Figure~\ref{fig:image_iw4nh}.

On the other hand, the right upper panel (B in Figure~\ref{fig:image_iw4nh} and Table~\ref{tab:iw4nh_and_para}), heavy obscuration is caused by dust, but the gas absorber does not cover the line-of-sight. 
Considering the X-ray emitting size is extremely compact with $<10$~$R_\mathrm{g}$ \citep[e.g.,][]{dai10,mor10} while optical and UV-emitting region of accretion disk is much larger, with $R \approx 100$--$10^3$~$R_\mathrm{g}$ \citep{kat08,mor10,ich19a},
these objects show unobscured properties with $\log (\NHunit)<22$ because of the lack of obscuring gas, but it still maintains high $(i-W4)_{\rm AB}$ values due to the significant dust extinction and dust re-emission.
This dust and gas geometry suggests a transition phase between an obscured AGN and an optically selected quasar in the gas-rich major merger scenario \citep[][]{2008ApJS..175..356H}. 
The majority of eFEDS-DOGs, with $\log (\NHunit)<22$ and $(i-W4)_{\mathrm{AB}}>7$ [mag], likely exhibit the geometry of gas and dust similar to (B) in Figure~\ref{fig:image_iw4nh}.

\subsection{X-ray and 6~$\mu$m Luminosity Correlation}\label{D_ER}

Our sample contains both of MIR and X-ray luminosity information of a statistically significant number of eFEDS-DOGs, and their locations in the luminosity-luminosity plane between MIR and X-ray enable us to discuss a rough indication of the Eddington ratio of the eFEDS-DOGs.
The tight correlation between X-ray and MIR luminosities for AGNs has been reported in numerous studies. The trend shows that almost 1:1 relation up to $\log (L_\mathrm{6 \mu m}/\mathrm{erg}~\mathrm{s}^{-1})\simeq \log (L_\mathrm{2-10~keV}~\mathrm{or}~L_\mathrm{0.5-2 keV}/\mathrm{erg}~\mathrm{s}^{-1})\simeq 44$ \citep[e.g.,][]{gan09,asm15,ich12,ich17,ich19,2012ApJ...753..104M,2015MNRAS.449.1422M,yam23}, above which X-ray luminosities show a saturation and thus the flatter trend \citep[e.g.,][]{2015ApJ...807..129S,2017ApJ...837..145C,2019MNRAS.484..196T,2022AandA...661A..15T,ich23}.
On the other hand, extremely IR luminous DOGs, or called hot-DOGs \citep[e.g.,][]{2012ApJ...755..173E,2021A&A...654A..37D}, are known to have more X-ray deficit feature compared to the MIR and X-ray relation discussed above, suggesting that hot-DOGs have higher Eddington ratio, possibly exceeding super-Eddington limit \citep{2014ApJ...794..102S,2016ApJ...819..111A,ric17,2018AandA...618A..28Z,2018MNRAS.474.4528V}.

\begin{figure}
   \centering
   \includegraphics[width=9cm]{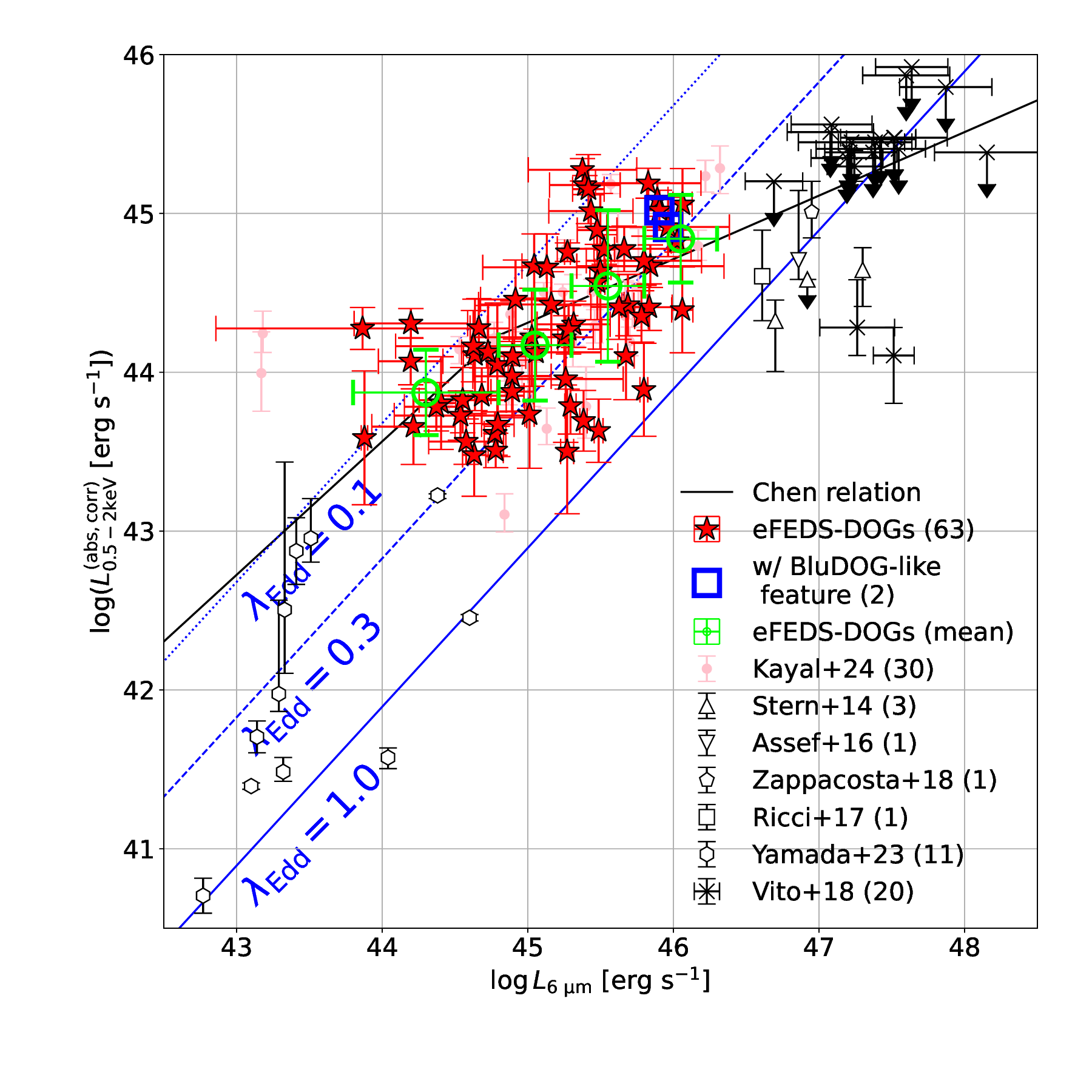}
   \caption{
   The luminosity correlation between the luminosities at absorption corrected 0.5--2~keV  ($\lxsoftabs$) and 6~$\mu$m ($\lsix$).
The symbols are the same as in Figure~\ref{fig:iw4}.
The green open circles show the mean and standard deviation of $\lxsoftabs$ for the eFEDS-DOG sample, and each bin has a range of $\log\lxsoftabs$ = 43.8--44.8, 44.8--45.3, 45.3--45.8, and 45.8--46.3. 
Open triangle, inverted triangle, square, pentagon, and black crosses represent
IR luminous quasars or hot-DOGs \citep{2012ApJ...755..173E,2021A&A...654A..37D} from \cite{2014ApJ...794..102S}, \cite{2016ApJ...819..111A}, \cite{ric17}, \cite{2018AandA...618A..28Z}, and \cite{2018MNRAS.474.4528V}, respectively.
Open hexagon represents a sample of \cite{yam23}; local ($z<0.1$) U/LIRGs with a final merger stage, with two nuclei in common envelope (stage = ``D'').
The black solid line represents the slope in the study of optically selected quasars at $1.5<z<3$, named Chen relation in this study \citep[][]{2015ApJ...807..129S,2017ApJ...837..145C}.
The blue solid, dashed, and dotted lines represent the expected location of Eddington ratio of $\eddington =  1.0, 0.3,$ and $0.1$, respectively, estimated by \cite{2019MNRAS.484..196T}.
Note that original Chen relation and $\lambda_{\mathrm{Edd}}$ lines are defined between $\lxhardabs$ and $\lsix$. We convert the original one to the $\lxsoftabs$--$\lsix$ relation by assuming $\Gamma=1.8$.
}
   \label{fig:L6L0520I}
\end{figure}

Figure~\ref{fig:L6L0520I} shows a luminosity correlation between the rest-frame 6~$\mu$m ($\lsix$) and $\lxsoftabs$ of our eFEDS-DOGs sample.
The expected curve of Eddington ratio of $\eddington = 0.1, 0.3,$ and $1.0$ are also over-plotted \citep{2019MNRAS.484..196T}. 
Our eFEDS-DOGs nicely covers the $L_\mathrm{6 \mu m}$ luminosity gap between high-$z$ ($z>2$) IR luminous hot DOGs at $\log (L_\mathrm{6 \mu m}/\mathrm{erg}~\mathrm{s}^{-1})>46.0$ and local U/LIRGs at $\log (L_\mathrm{6 \mu m}/\mathrm{erg}~\mathrm{s}^{-1})<44.5$ \citep{yam23}.
Overall, most eFEDS-DOGs follow the trend of the luminosity relation by \cite{2017ApJ...837..145C} (hereafter, we referred to as Chen relation).
This suggests that eFEDS-DOGs on average do not show extremely high Eddington ratio feature unlike the hot-DOGs, but rather they would have a similar Eddington ratio with quasars  ($\left<\eddington \right> \approx 0.3$) with similar MIR luminosity range \citep[e.g.,][]{kol06}.
Similarly, the sources from \cite{2024MNRAS.531..830K} exhibit a comparable distribution in Figure~\ref{fig:L6L0520I}, overlapping well with our eFEDS-DOGs sample. This further supports the idea that DOGs in general tend to populate this parameter space, as also indicated by the findings of \cite{2024MNRAS.531..830K}.
However, focusing the individual sources, the scatter of eFEDS-DOGs in Figure~\ref{fig:L6L0520I} is huge with $\eddington$ standard deviation $\sigma=0.45$~dex, and some eFEDS-DOGs deviate over 0.5~dex above/below the Chen relation. Since we use a simple X-ray spectral fitting with average $\log (\NHunit)$ uncertainly of 0.69~dex as discussed in Section~2.1.1., this may partially contribute to the observed scatter. Beyond this methodological contribution, we discuss possible physical origins of these scatters.

One possible origin of this deviation is the difference in Eddington ratio for eFEDS-DOGs, especially for the sources below the Chen relation.
The eFEDS-DOGs at $\sim$1 dex below the Chen relation align closely with the $\eddington=1$ line around $\lsix\sim10^{45.5}$ erg s$^{-1}$, suggesting that they are in the Eddington-limit phase (or possibly in the super-Eddington). 
Objects with high Eddington ratios may be undergoing more rapid mass accretion \citep[known as the Slim disk model;][]{1988ApJ...332..646A}. 
In this model, when the mass accretion rate significantly exceeds the Eddington rate, radiation becomes trapped within the accreting matter before escaping by radiative diffusion, resulting in a radiatively inefficient state \citep[e.g.,][]{2009PASJ...61L...7O,2005ApJ...628..368O,2014MNRAS.441.3177M,2016MNRAS.459.3738I,2020ARA&A..58...27I,2020MNRAS.497..302T}.
According to this model, the inner region (X-ray emission region) of the accretion disk is expected to be thicker than the outer accretion disk {\citep[e.g.,][]{2009PASJ...61L...7O}.}
Consequently, photons from the accretion disk may not reach the inner region due to the X-ray-emitting region being covered by a thick disk,
resulting in being observed as X-ray weak AGNs, as the reduced photon supply into the X-ray-emitting region diminishes their observed luminosity. 
A similar trend was also reported in several hot DOGs and X-ray weak AGNs with super-Eddington signature (see Figure~\ref{fig:L6L0520I}), showing extremely X-ray weak feature \citep[e.g.,][]{2014ApJ...794..102S,2016ApJ...819..111A,ric17,2018MNRAS.474.4528V,2018AandA...618A..28Z,yam23}
as well as recently reported ``little red dots'' discovered by JWST with more extreme X-ray ``deficit'' feature \citep{koc24,yue24}.
The eFEDS-DOGs at $\sim$1 dex below the Chen relation might be low-$z$ ($z\sim1$) and low $\lsix$ analogous sources of such hot DOGs sample found mostly at $z>2$.

Figure~\ref{fig:L6L0520I} also shows sources above the Chen relation, up to $+$0.5 dex.
The origins of such deviation have three possibilities:
1) over-estimation of photo-{\textit{z}}, 2) objects with a jet, 3) lensed objects.
The first possibility is that the photo-$z$ values of those sources might be over-estimated. The over-estimation of photo-$z$ causes the boost of both luminosities by holding the 1:1 relation. In this case, such sources deviate from the Chen relation that has the shallower slope in the high luminosity end.
Two eFEDS-DOGs locating above $+$0.5 dex from the Chen relation exhibits
$\lxsoftabs\sim 10^{45}$~erg~s$^{-1}$ (see Figure~\ref{fig:L6L0520I}), with objects located around $z\sim1.5$ (see Figure~\ref{fig:LXz_w_os}). 
Assuming that the corrected redshifts have the median value of eFEDS-DOGs sample of $z\sim1$, 
the boost factor $\mu$ of both luminosities from $z\sim1$ to $1.5$ is $\mu \approx 3$, mitigating the gap for most of the sources.
However, since we selected the eFEDS clean sample with CTP\_REDSHIFT\_GRADE $\geq$ 3 (see Section~\ref{SSS_CS}), the likelihood of overestimating the photo-$z$ of eFEDS-DOGs appears to be low.

The second possibility is that those sources have prominent jet emission, or radio-loud eFEDS-DOGs. The jet emission of those sources can contaminate to X-ray bands, boosting the observed X-ray emission \citep[e.g., Blazars and some radio-loud AGN;][]{mil11,wu13,2017MNRAS.469..255G,zhu20}. 
In \cite{2012ApJ...753..104M}, blazars and radio-loud AGN exhibit a 1:1 relation between $L_{\rm X}$ and $L_{\rm MIR}$, suggesting that the sub-sample of eFEDS-DOGs at $+$0.5~dex from the Chen relation may have radio jets, which should be luminous in the radio wavelength.
However, this scenario would be unlikely, since 
only one source at least 0.5~dex above the Chen relation is detected in the VLA/FIRST 1.5~GHz \citep{bec95,whi97,hel15} or VLASS 3~GHz radio bands \citep{lac20}. 
In addition, the median value of the X-ray photon index of this sources is $\Gamma =1.95$. 
Thus they do not show the characteristic feature of radio-jet in X-ray spectra, which tend to show a shallower photon index, with $\left<\Gamma\right>=1.4$ \citep[e.g.,][]{igh19}.

The final possibility is the scenario that the emission of those sources are gravitationally lensed, boosting the both MIR and X-ray luminosities in a 1:1 relation \citep[e.g.,][]{con21,2023MNRAS.tmp.2823C}. 
Assuming a tenfold increase in luminosity for the sub-sample of eFEDS-DOGs at $+$0.5 dex, they align with the Chen relation. The magnification factor can reach $\eta=10$--$500$ \citep{gli18,fan19,fuj20}, thus the de-lensed luminosities of those sources can join the 1:1 relation in the luminosity-luminosity plane.

In summary, at present, we cannot conclude which of the three scenarios are preferable.
Additional follow-up observations such as the optical spectroscopy to determine the spec-$z$ (for scenario 1), deeper radio observations (for scenario 2), and/or high spatial resolution imaging (for scenario 3) are necessary to narrow down those three origins.

\subsection{Eddington ratio and other properties of eFEDS-DOGs}\label{D_ERED}

The location of sources in Figure~\ref{fig:L6L0520I} 
allows us to estimate $\lambda_{\mathrm{Edd}}$, which enable us to compare $\eddington$ with other parameters obtained by X-ray observations.
Several observational X-ray spectral studies of AGN and quasars indicate that photon index $\Gamma$
changes as a function of $\lambda_{\mathrm{Edd}}$ \citep[see ][]{1996A&A...305...53B,she08,ris09,2013MNRAS.433.2485B, tra17}, which is thought to originate from the change of the coronal properties around the accretion disk, specially a plasma parameter ($kT_e$) and a plasma optical depth ($\tau$), where $k$ and $T_e$ are the Boltzmann constant and electron temperature, respectively \citep[][]{1995ApJ...450..876T,1996MNRAS.283..193Z,2010A&A...512A..58I}.

\begin{figure}
   \includegraphics[width=9cm]{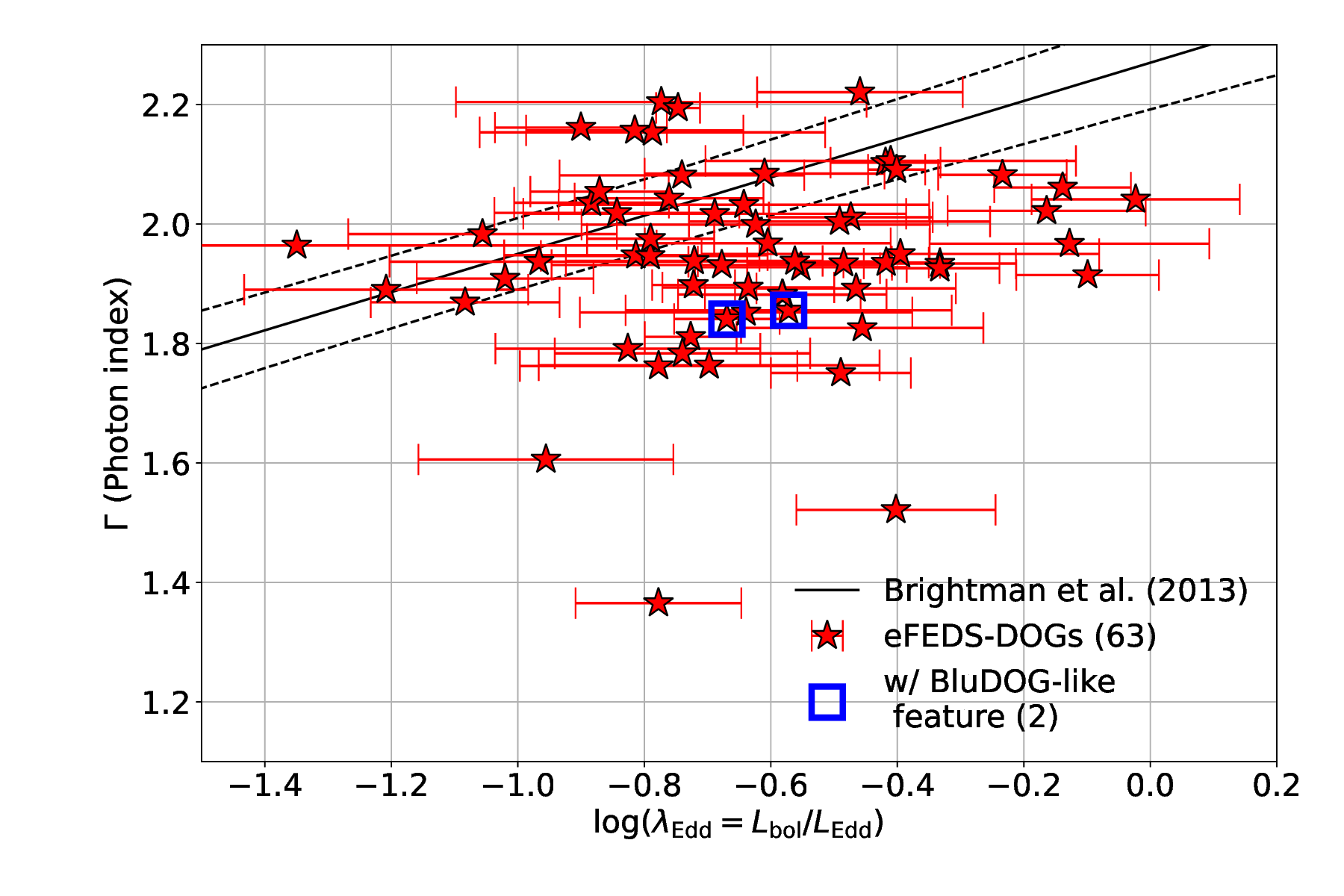}
   \caption{The $\Gamma$--$\lambda_{\mathrm{Edd}}$ diagram. The symbols are the same as in Figure~\ref{fig:iw4}. Black solid and dashed lines represent the $\Gamma$--$\lambda_{\mathrm{Edd}}$ relation and $\pm1\sigma$ lines from $\Gamma$--$\lambda_{\mathrm{Edd}}$ relation \citep[][]{2013MNRAS.433.2485B}.}
   \label{fig:gammalambda}
\end{figure}

\begin{figure}
   \includegraphics[width=9cm]{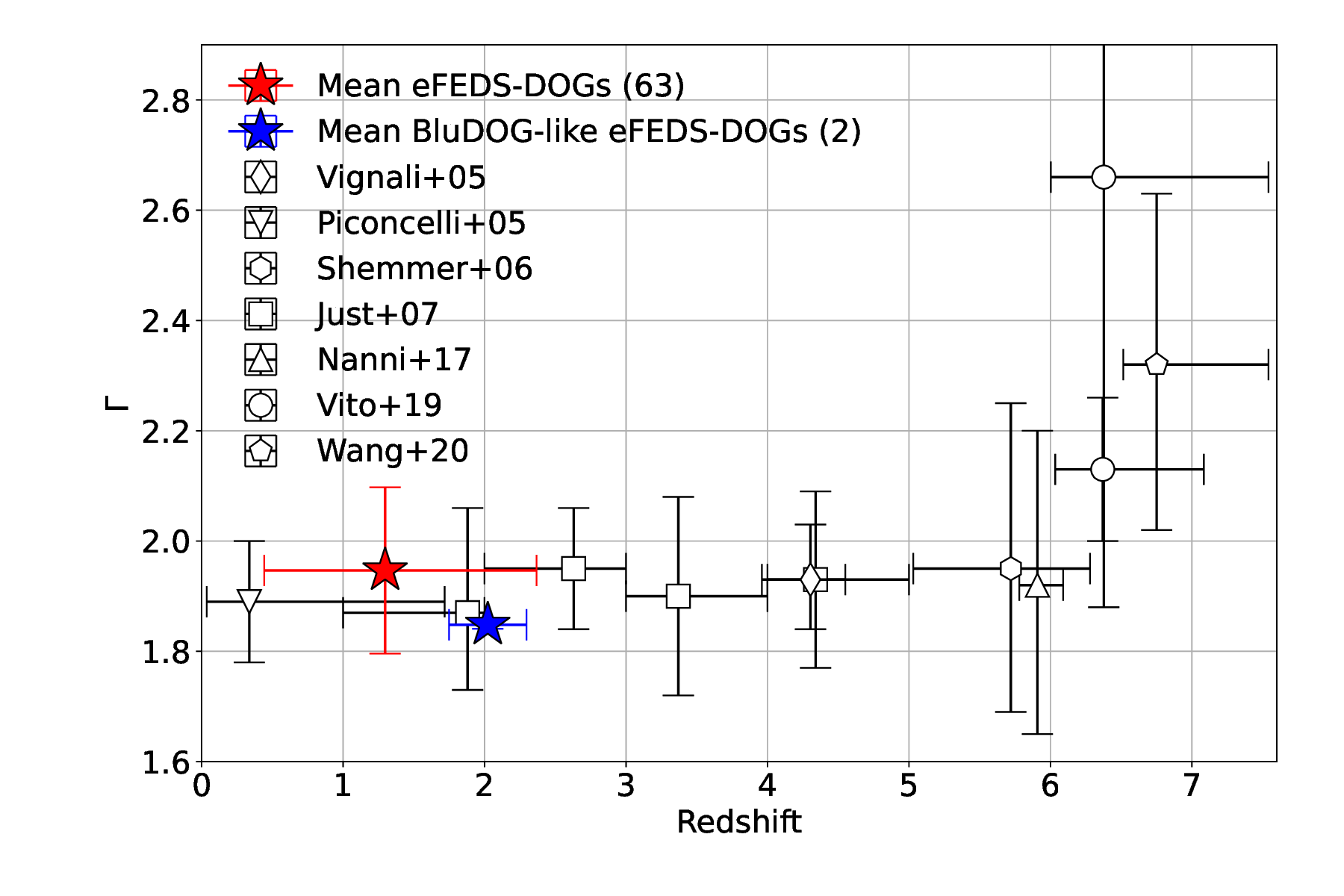}
   \caption{The $\Gamma$--$z$ diagram. The symbols are the same as in Figure~\ref{fig:iw4}. Open diamond, inverted triangle, hexagon, square, triangle, circle, and pentagon plots denote X-ray detected quasars from \cite{2005AJ....129.2519V}, \cite{2005A&A...432...15P}, \cite{2006ApJ...644...86S}, \cite{2007ApJ...665.1004J}, \cite{2017A&A...603A.128N}, \cite{2019A&A...630A.118V}, and \cite{2021ApJ...908...53W}, respectively. }
   \label{fig:gammaz}
\end{figure}

Figure~\ref{fig:gammalambda} shows the $\Gamma$--$\lambda_{\mathrm{Edd}}$ relation of eFEDS-DOGs, with $\Gamma$ and $\lambda_{\mathrm{Edd}}$ of eFEDS-DOGs being $\Gamma = 1.95\pm0.15$ and $\log\lambda_{\mathrm{Edd}} = -0.64\pm0.27$, respectively. Although the median value of eFEDS-DOGs is slightly below the $\Gamma$--$\lambda_{\mathrm{Edd}}$ relation obtained by \cite{2013MNRAS.433.2485B}, it still falls within the uncertainty range, considering the large uncertainties in the $\eddington$ estimation used here. This trend also persists when comparing eFEDS-DOGs with other X-ray detected quasars in the plane of $\Gamma$ as a function of redshift, as illustrated in Figure~\ref{fig:gammaz}. Our averaged values are largely consistent with previous studies at $z\sim1$--$2$, suggesting that the coronal properties of eFEDS-DOGs are similar to optically selected quasars at similar redshifts.

\begin{figure}
   \includegraphics[width=1.0\linewidth]{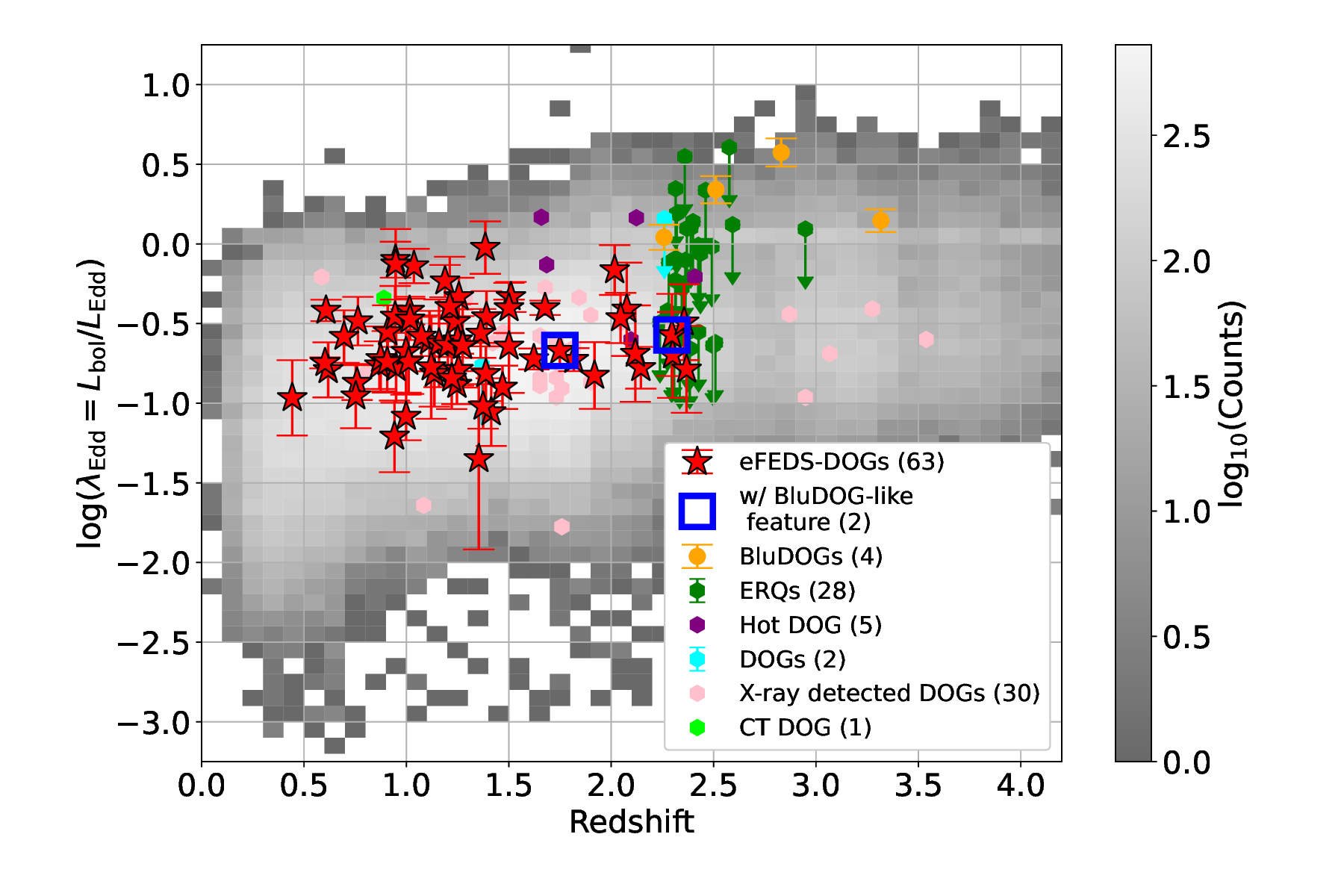}
   \caption{$\log\lambda_{\mathrm{Edd}}$ as a function of redshift. The gray 2D histogram represents the number density of SDSS quasars \citep[][]{2011ApJS..194...45S}. Red stars and blue squares denote eFEDS-DOGs and BluDOG-like eFEDS-DOGs, respectively. Orange, green, purple, cyan, pink, and lime green hexagons denote BluDOGs \citep[][]{2022ApJ...941..195N}, extremely red quasars \citep[ERQs:][]{2019MNRAS.488.4126P}, hot DOGs \citep[][]{2018ApJ...852...96W}, DOGs \citep[][]{2011AJ....141..141M}, X-ray detected DOGs \citep[][]{2024MNRAS.531..830K}, and CT DOG \citep[][]{2020ApJ...888....8T}, respectively.}
   \label{fig:lambdaz}
\end{figure}

Figure~\ref{fig:lambdaz} shows $\eddington$ of eFEDS-DOGs as a function of redshift. The underlying gray contour represents the number density of SDSS quasars in the plane \citep{2011ApJS..194...45S}.
Note that the $\log\lambda_{\mathrm{Edd}}$ values for the X-ray detected DOGs from \cite{2024MNRAS.531..830K} are not directly taken from their study but are instead estimated using the method of \cite{2019MNRAS.484..196T} based on $\lsix$ and $\lxsoftabs$. This approach was adopted to maintain consistency with our analysis.
The $\log\lambda_{\mathrm{Edd}}$ values of SDSS quasars, eFEDS-DOGs, and BluDOG-like eFEDS-DOGs at each redshift are summarized in Table~\ref{tab:lambdaz_dist}.
At each redshift range, the averaged $\eddington$ of eFEDS-DOGs are higher than those of SDSS quasars,
but they are again within the scatter each other.
This result supports that our eFEDS-DOGs sample has similar level or possibly more rapidly growing SMBHs compared to SDSS quasars at each redshift, but roughly half of them are behind the curtain of obscuring gas of $\log(\NHunit)>22$.

\begin{figure}
   \includegraphics[width=9cm]{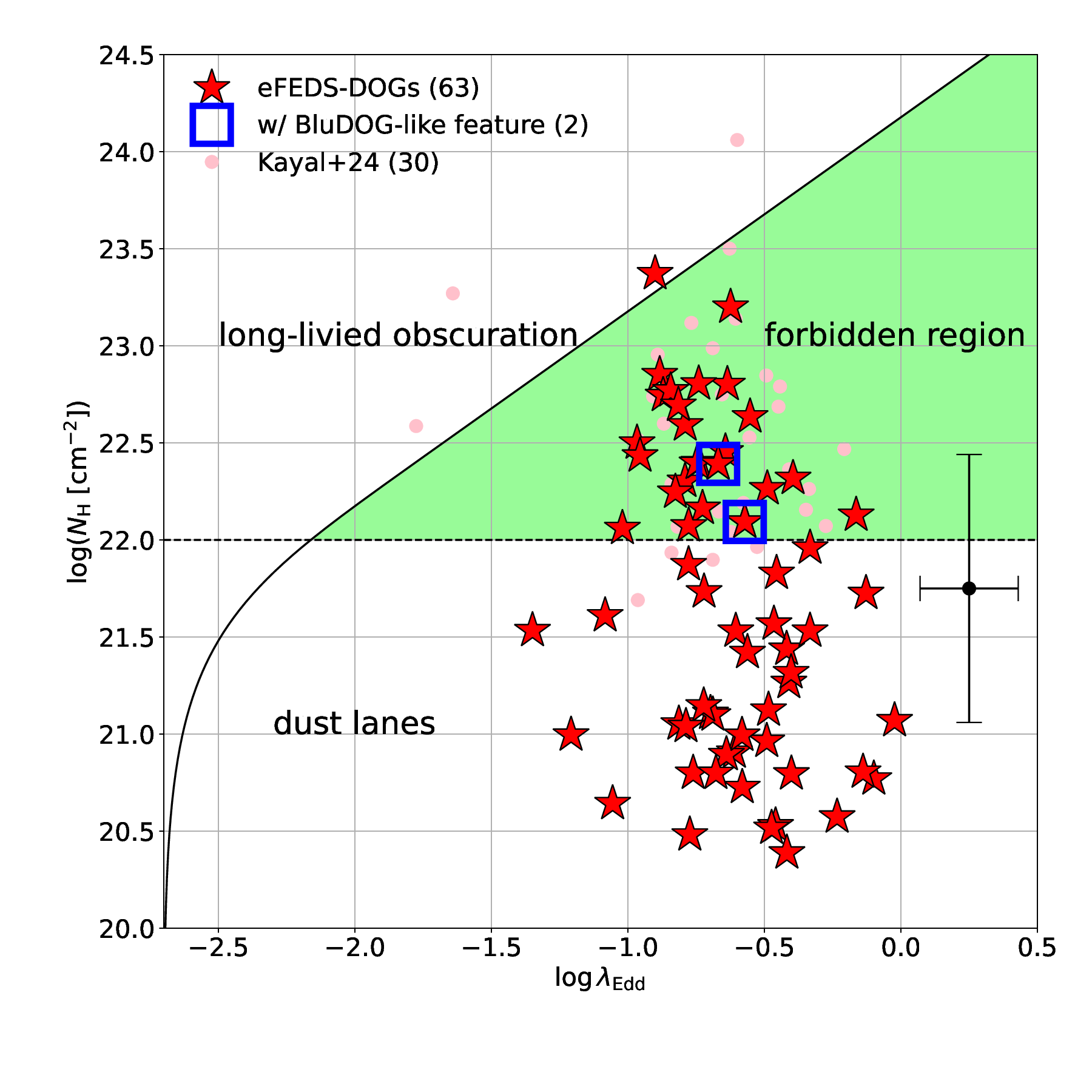}
   \vspace{-1cm}
   \caption{The $\NH$--$\lambda_{\mathrm{Edd}}$ diagram. The symbols are the same as in Figure~\ref{fig:iw4}. A gray plot with error bars denotes typical standard deviations of $\NH$ and $\lambda_{\mathrm{Edd}}$. The standard deviations of $\NH$ and $\lambda_{\mathrm{Edd}}$ are 0.69 dex and 0.18 dex, respectively. Black dashed and solid lines denote $\log(N_{\mathrm{H}}\ {\mathrm{[cm^{-2}]}})=22.0$ and equation (6) in \cite{2018MNRAS.479.3335I}. Green filled region shows a forbidden region.}
   \label{fig:lambdaNH}
\end{figure}

The results presented above indicate that eFEDS DOGs exhibit similar X-ray coronal and accretion disk properties to optically bright quasars within a similar redshift range. However, the central engine of eFEDS DOGs remains obscured by dust and gas. This dust and gas-rich environment surrounding the central engine positions eFEDS DOGs uniquely in the plane of $\NH$ vs $\eddington$, as shown in Figure~\ref{fig:lambdaNH}. The black solid line represents the effective Eddington limit \citep[see also][]{2006MNRAS.373L..16F,2009MNRAS.394L..89F,2017Natur.549..488R} derived from equation (6) in \cite{2018MNRAS.479.3335I}, assuming IR and UV opacities ($\kappa_{\mathrm{IR}}$ and $\kappa_{\mathrm{UV}}$) of $\kappa_{\mathrm{IR}}=0$ and $\kappa_{\mathrm{UV}}=200$ ${\mathrm{cm^2\ g^{-1}}}$.

In this plane, long-live clouds can survive against radiation pressure (``long-lived obscuration'' in Figure~\ref{fig:lambdaNH}), while absorbed objects at low column densities may appear above their effective Eddington limit due to the presence of dust lanes in the galaxy (``dust lanes'' in Figure~\ref{fig:lambdaNH}).
AGNs in ``forbidden region'' are expected to have short-lived absorption due to significant radiation pressure on the absorbing clouds.
Therefore AGNs do not stay in this region for prolonged periods, suggesting that AGNs are in a ``blowout'' phase.
Out of 63 eFEDS-DOGs, 25 sources (including two bluDOG-like eFEDS-DOGs) are in the forbidden region.
Given that 23 objects from the \cite{2024MNRAS.531..830K} DOGs sample are also located in the forbidden region, the presence of nearly half of the eFEDS-DOGs in this region further supports the expectation that X-ray detected DOGs tend to be in blow-out phase.
This result supports the idea that eFEDS-DOGs in the forbidden region experience the gas outflow of the obscuring material, which would eventually lead to a decrease in $\NH$. Consequently, the sources would be preferentially observed as unobscured AGN \citep{ric22,ric23}. Our results indicate that unobscured eFEDS DOGs would be such sources. Thus, eFEDS-DOGs are prominent candidates in a previously missed transitional phase between obscured AGN and optically selected quasars
(see also Section~\ref{D_NH}).

\begin{table*} 
\caption{$\log\lambda_\mathrm{Edd}$ distribution at each redshift range}  
\label{tab:lambdaz_dist}      
\centering          
\begin{tabular}{c c c c c c c}
\hline\hline  
                        & SDSS quasars  &       & eFEDS-DOGs &      & BluDOG-like eFEDS-DOGs & \\
\textbf{Redshift range} & $\log\lambdaedd$ & Counts & $\log\lambdaedd$  & Counts & $\log\lambdaedd$ & Counts \\\hline
0.0--0.5 & $-1.07\pm0.50$ &(8490)   & $-0.97$        & (1)  & nan               & (0)\\
0.5--1.0 & $-0.92\pm0.41$ &(19130)  & $-0.67\pm0.28$ & (18) & nan               & (0)\\
1.0--1.5 & $-0.81\pm0.34$ &(25586)  & $-0.64\pm0.29$ & (27) & nan               & (0)\\
1.5--2.0 & $-0.73\pm0.37$ &(28031)  & $-0.59\pm0.17$ & (8)  & $-0.67$           & (1)\\
2.0--2.5 & $-0.60\pm0.38$ &(12345)  & $-0.56\pm0.19$ & (9)  & $-0.57$           & (1)\\\hline\hline
0.0--2.5 & $-0.80\pm0.41$ &(93582) & $-0.63\pm0.26$ & (63) & $-0.62\pm0.05$    & (2)\\\hline
\end{tabular}
\end{table*}

\subsection{Lifetime of eFEDS-DOGs}\label{D_LT}

eROSITA revealed newly X-ray detected DOGs, which constitutes only 1.3\% (74/5738) out of the total population, but this population might be in a transient phase from highly obscured AGN to unobscured ones.
This rarity would give us important information on the lifetime of such a transient phase.

The lifetime of HSC-WISE DOGs at $z\sim1$ is expected to be approximately 40 Myr, as suggested by DOG duty cicle \citep[][]{2017ApJ...835...36T}.
Assuming that eFEDS-DOGs represent one stage in the gas-rich major merger scenario, the lifetime of eFEDS-DOGs is naively estimated to be $\sim0.5$~Myr, given that the abundance of eFEDS-DOGs among all DOGs is 1.3\% (74/5738). 
This number fraction is similar to that of BluDOGs in DOGs \citep{2019ApJ...876..132N}. 
It is important to note that eFEDS-undetected DOGs may include objects such as eFEDS-DOGs viewed from the edge-on angle. 
Therefore, the estimated $0.5$~Myr should be treated as the lower-bound for an early transition phase between an obscured AGN and an optically thin quasar (see Section~\ref{D_NH}).

\subsection{BluDOG-like eFEDS-DOGs}\label{D_BL}

Blue-excess objects in dusty AGNs are crucial because they are thought to be in a dusty outflow phase \citep[][]{2019ApJ...876..132N}, and BluDOG spectra exhibit a blue tail on the C~{\sc iv} emission line \citep[][]{2022ApJ...941..195N}, suggesting that the blue tail is evidence of nucleus outflow. 
Additionally, some BluDOGs show broad emission lines that enable us to estimate the black hole mass and thus the Eddington ratio \citep{2022ApJ...941..195N}. 
Their Eddington ratio show $\eddington \gtrsim 1$, suggesting that they are in a super-Eddington accretion phase. 
Recently, \cite{2023ApJ...959L..14N} identified high-$z$ BluDOG candidates from JWST-ERO high-$z$ galaxy samples at $5<z<7$ \citep[e.g.,][]{2023ApJ...956...61A,2023arXiv230514418B,2023arXiv230805735F,2023ApJ...952..142F,2023ApJ...954L...4K,2023ApJ...957L...7K,2023arXiv230607320L,2023arXiv230605448M}. 
The results imply that the blue-excess DOGs-like objects ubiquitously exist, at least over the cosmic time at $2<z<7$. 
Thus, the finding of lower-$z$ analogs at $z<2$ would be of special interest as candidates of AGN achieving super-Eddington accretion.

\begin{figure}
   \centering
   \includegraphics[width=8.5cm]{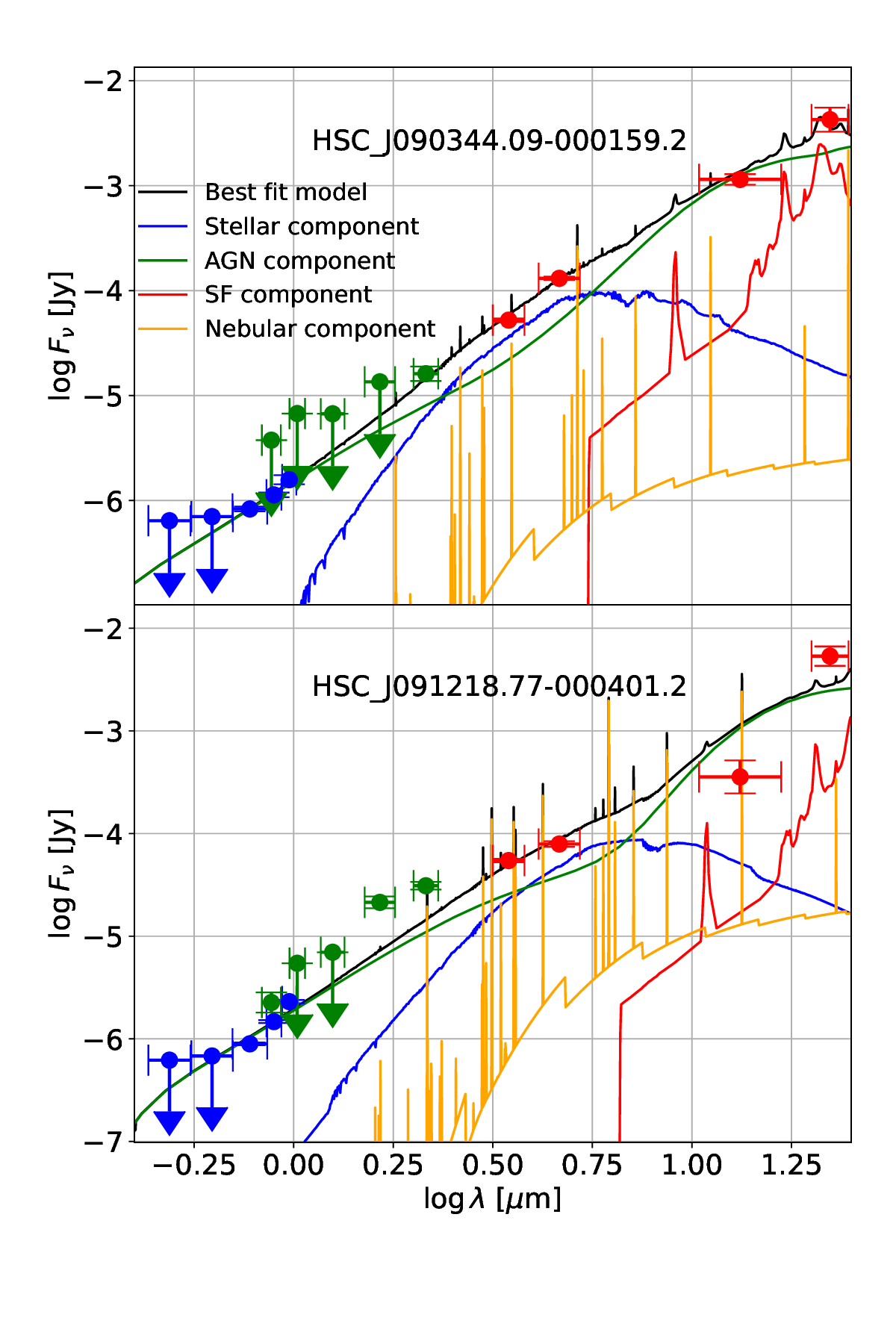}
   \caption{SEDs of BluDOG-like objects. The pionts and lines are same as in Figure~\ref{fig:exampleSED}.  Arrows denote upper limits.}
   \label{fig:BluLIKESED}
\end{figure}

Unfortunately, none of our eFEDS-DOGs samples satisfies the conventional BluDOG criteria of $\alpha_{grizy}^{\rm opt}<0.4$, but this criterion is optimized for sources whose C~{\sc{iv}} emission lines are in the traditional optical range at $5000 < \lambda_{\mathrm{obs}} < 7000$~\AA, therefore at $2.2<z<3.5$.
On the other hand, two eFEDS-DOGs show $\alpha_{gri}^{\rm opt} < 0.8$ (Section \ref{OPTICAL_CLASS_DOGs}), and this blue excess would trace either the AGN continuum and/or the strong {Mg~\sc{ii}}$\lambda2800$ emission line.
Figure~\ref{fig:BluLIKESED} shows the SEDs of BluDOG-like objects.
The SEDs (1--20\ $\mu$m) of HSC J090344.09$-$000159.2 and HSC J091218.77$-$000401.2 exhibit power-law emission, and the optical SEDs do not show extinction. 
These characteristics are similar to those of BluDOGs found by \cite{2019ApJ...876..132N}. 
We suggest that one reason why we cannot select BluDOG-like objects using $\alpha^{\rm opt}_{grizy} < 0.4$ is the difference in redshift ranges between BluDOGs and eFEDS-DOGs. 
In \cite{2022ApJ...941..195N}, the redshift range of BluDOGs selected by $\alpha^{\rm opt}_{grizy} < 0.4$ is $2.2 < z_{\rm spec} < 3.3$, while the photo-$z$ of the BluDOG-like objects is 1.749 and 2.296. 
In Figure~\ref{fig:BluLIKESED}, the HSC $z$- and $y$-bands show flux excess compared with that of HSC $g$-, $r$-, and $i$-bands.

\section{Conclusions}\label{C}
We have constructed the sample of 74 (65) X-ray detected dust obscured galaxies (DOGs) detected in the 0.5--2~keV bands by the eROSITA/eFEDS survey, and we call them ``eFEDS detected DOGs''. Among them, 65 sources have secured X-ray spectral fitting results, enable us to investigate X-ray properties with statistically large sample, and we call them eFEDS-DOGs. The eFEDS-DOGs were identified by the combination of multi-wavelength dataset of HSC and LS8 optical, VIKING NIR, unWISE MIR bands, and eROSITA eFEDS 0.5--2~keV bands.
Our results are summarized as follows.

\begin{enumerate}
     \item eFEDS-DOGs are located in a unique region characterized by $i_\mathrm{AB}=22$--$24$~mag and $m_\mathrm{22 \mu m}=14$--$16$~mag, which were unexplored by previous X-ray studies of DOGs.
     \item The most of DOGs are not detected by \textit{eROSITA}. By assuming that they would follow the fiducial $\lxhardabs$--$\lsix$ relation by \cite{2017ApJ...837..145C} and the fiducial photon index of $\Gamma=1.8$, expected lower-bound of $\NH$ reaches $\NH \gtrsim 10^{23}$ cm$^{-2}$, suggesting that the most of DOGs are heavily obscured by gas. 
     \item A large fraction of eFEDS-DOGs are not heavily obscured by gas due to $\NH \lesssim 10^{22}$ cm$^{-2}$. Therefore, eFEDS-DOGs contain gas+dust unobscured DOGs.
     \item Examination of the $L_{\rm 6\ \mu m}$ vs. $\lxsoftabs$ diagram reveals a predominant alignment with the $L_{\rm 6\ \mu m}$--$L_{\rm X}$ relation. However, some eFEDS-DOGs exhibit deviations, down to $\simeq1.0$~dex below the $L_{\rm 6\ \mu m}$--$L_{\rm X}$ relation. The sources located $-$1 dex below the relation may signal high Eddington ratios reaching the Eddington limit.
     \item 25 out of 63 eFEDS-DOGs are located in the forbidden region on the $\NH$ vs. $\eddington$ plane, supporting the idea that eFEDS-DOGs in the forbidden region experience gas outflow of the obscuring material, which would eventually lead to a decrease in $\NH$.
     eFEDS-DOGs are prominent candidates in a previously missed transitional phase between obscured AGN and optically selected quasars.
     \item Despite none of the eFEDS-DOGs meeting the strict BluDOG criteria, our identification of BluDOG-like objects based on $\alpha^{\mathrm{opt}}_{gri} < 0.8$ and MIR\_CLASS = 2 (PL DOGs) unveils a compelling subset. Their spectral energy distributions (SEDs) exhibit a power-law trend between the optical and MIR regions, coupled with a distinct flattening between HSC $g$-band and HSC $i$-band. This suggests that these BluDOG-like objects may represent a population of low-redshift BluDOGs.
\end{enumerate}

\begin{acknowledgements}
The authors gratefully acknowledge the anonymous referee for a careful reading of the manuscript and very helpful comments.
We are honored and grateful for the opportunity to observe the universe from Mauna Kea, which has cultural, historical, and natural significance in Hawaii. 
The Hyper Suprime-Cam (HSC) Collaboration includes the astronomical communities of Japan, Taiwan, and Princeton University. 
The HSC instrumentation and software were developed by the National Astronomical Observatory of Japan (NAOJ), the Kavli Institute for the Physics and Mathematics of the Universe (Kavli IPMU), the University of Tokyo, the High Energy Accelerator Research Organization (KEK), the Academia Sinica Institute for Astronomy and Astrophysics in Taiwan (ASIAA), and Princeton University. Funding was contributed by the FIRST program from the Japanese Cabinet Office, the Ministry of Education, Culture, Sports, Science and Technology (MEXT), the Japan Society for the Promotion of Science (JSPS), Japan Science and Technology Agency (JST), the Toray Science Foundation, NAOJ, Kavli IPMU, KEK, ASIAA, and Princeton University. 
This paper makes use of software developed for the Large Synoptic Survey Telescope. 
We thank the LSST Project for making their code available as free software at http://dm.lsstcorp.org.
The Pan-STARRS1 Surveys (PS1) have been made possible through contributions of the Institute for Astronomy, the University of Hawaii, the Pan-STARRS Project Office, the Max-Planck Society and its participating institutes, the Max Planck Institute for Astronomy, Heidelberg and the Max Planck Institute for Extraterrestrial Physics, Garching, The Johns Hopkins University, Durham University, the University of Edinburgh, Queen's University Belfast, the Harvard-Smithsonian Center for Astrophysics, the Las Cumbres Observatory Global Telescope Network Incorporated, the National Central University of Taiwan, the Space Telescope Science Institute, the National Aeronautics and Space Administration under Grant No. NNX08AR22G issued through the Planetary Science Division of the NASA Science Mission Directorate, the National Science Foundation under Grant No. AST-1238877, the University of Maryland, and Eotvos Lorand University (ELTE).
This publication has made use of data from the VIKING survey from VISTA at the ESO Paranal Observatory, program ID 179.A-2004.
Data processing has been contributed by the VISTA Data Flow System at CASU, Cambridge and WFAU, Edinburgh. 
This publication makes use of data products from the Wide-field Infrared Survey Explorer, which is a joint project of the University of California, Los Angeles, and the Jet Propulsion Laboratory/California Institute of Technology, funded by the National Aeronautics and Space Administration. 
Herschel is an ESA space observatory with science instruments provided by Europeanled Principal Investigator consortia and with important participation from NASA. 
The Legacy Surveys consist of three individual and complementary projects: the Dark Energy Camera Legacy Survey (DECaLS; Proposal ID \#2014B-0404; PIs: David Schlegel and Arjun Dey), the Beijing-Arizona Sky Survey (BASS; NOAO Prop. ID \#2015A-0801; PIs: Zhou Xu and Xiaohui Fan), and the Mayall z-band Legacy Survey (MzLS; Prop. ID \#2016A-0453; PI: Arjun Dey). 
DECaLS, BASS and MzLS together include data obtained, respectively, at the Blanco telescope, Cerro Tololo Inter-American Observatory, NSF’s NOIRLab; the Bok telescope, Steward Observatory, University of Arizona; and the Mayall telescope, Kitt Peak National Observatory, NOIRLab. 
Pipeline processing and analyses of the data were supported by NOIRLab and the Lawrence Berkeley National Laboratory (LBNL). 
The Legacy Surveys project is honored to be permitted to conduct astronomical research on Iolkam Du’ag (Kitt Peak), a mountain with particular significance to the Tohono O’odham Nation.
NOIRLab is operated by the Association of Universities for Research in Astronomy (AURA) under a cooperative agreement with the National Science Foundation. 
LBNL is managed by the Regents of the University of California under contract to the U.S. Department of Energy.
This project used data obtained with the Dark Energy Camera (DECam), which was constructed by the Dark Energy Survey (DES) collaboration. 
Funding for the DES Projects has been provided by the U.S. Department of Energy, the U.S. National Science Foundation, the Ministry of Science and Education of Spain, the Science and Technology Facilities Council of the United Kingdom, the Higher Education Funding Council for England, the National Center for Supercomputing Applications at the University of Illinois at Urbana-Champaign, the Kavli Institute of Cosmological Physics at the University of Chicago, Center for Cosmology and Astro-Particle Physics at the Ohio State University, the Mitchell Institute for Fundamental Physics and Astronomy at Texas A\&M University, Financiadora de Estudos e Projetos, Fundacao Carlos Chagas Filho de Amparo, Financiadora de Estudos e Projetos, Fundacao Carlos Chagas Filho de Amparo a Pesquisa do Estado do Rio de Janeiro, Conselho Nacional de Desenvolvimento Cientifico e Tecnologico and the Ministerio da Ciencia, Tecnologia e Inovacao, the Deutsche Forschungsgemeinschaft and the Collaborating Institutions in the Dark Energy Survey. The Collaborating Institutions are Argonne National Laboratory, the University of California at Santa Cruz, the University of Cambridge, Centro de Investigaciones Energeticas, Medioambientales y Tecnologicas-Madrid, the University of Chicago, University College London, the DES-Brazil Consortium, the University of Edinburgh, the Eidgenossische Technische Hochschule (ETH) Zurich, Fermi National Accelerator Laboratory, the University of Illinois at Urbana-Champaign, the Institut de Ciencies de l’Espai (IEEC/CSIC), the Institut de Fisica d’Altes Energies, Lawrence Berkeley National Laboratory, the Ludwig Maximilians Universitat Munchen and the associated Excellence Cluster Universe, the University of Michigan, NSF’s NOIRLab, the University of Nottingham, the Ohio State University, the University of Pennsylvania, the University of Portsmouth, SLAC National Accelerator Laboratory, Stanford University, the University of Sussex, and Texas A\&M University.
BASS is a key project of the Telescope Access Program (TAP), which has been funded by the National Astronomical Observatories of China, the Chinese Academy of Sciences (the Strategic Priority Research Program “The Emergence of Cosmological Structures” Grant \# XDB09000000), and the Special Fund for Astronomy from the Ministry of Finance. The BASS is also supported by the External Cooperation Program of Chinese Academy of Sciences (Grant \# 114A11KYSB20160057), and Chinese National Natural Science Foundation (Grant \# 12120101003, \# 11433005).
The Legacy Survey team makes use of data products from the Near-Earth Object Wide-field Infrared Survey Explorer (NEOWISE), which is a project of the Jet Propulsion Laboratory/California Institute of Technology. NEOWISE is funded by the National Aeronautics and Space Administration.
The Legacy Surveys imaging of the DESI footprint is supported by the Director, Office of Science, Office of High Energy Physics of the U.S. Department of Energy under Contract No. DE-AC02-05CH1123, by the National Energy Research Scientific Computing Center, a DOE Office of Science User Facility under the same contract; and by the U.S. National Science Foundation, Division of Astronomical Sciences under Contract No. AST-0950945 to NOAO.
This work is based on data from eROSITA, the soft X-ray instrument aboard SRG, a joint Russian-German science mission supported by the Russian Space Agency (Roskosmos), in the interests of the Russian Academy of Sciences represented by its Space Research Institute (IKI), and the Deutsches Zentrum f\"{u}r Luftund Raumfahrt (DLR). The SRG spacecraft was built by Lavochkin Association (NPOL) and its subcontractors, and is operated by NPOL with support from the Max Planck Institute for Extraterrestrial Physics (MPE). The development and construction of the eROSITA X-ray instrument was led by MPE, with contributions from the Dr. Karl Remeis Observatory Bamberg \& ECAP (FAU Erlangen-Nuernberg), the University of Hamburg Observatory, the Leibniz Institute for Astrophysics Potsdam (AIP), and the Institute for Astronomy and Astrophysics of the University of T\"{u}bingen, with the support of DLR and the Max Planck Society. The Argelander Institute for Astronomy of the University of Bonn and the Ludwig Maximilians Universit\"{a}t Munich also participated in the science preparation for eROSITA.
This research made use of Astropy,\footnote{\url{https://www.astropy.org}} a community-developed core Python package for Astronomy \citep[][]{2013A&A...558A..33A, 2018AJ....156..123A,2022ApJ...935..167A}.
This study was financially supported by the Japan Society for the Promotion of Science (JSPS) KAKENHI grant No.~20H01939 (K.I.).

\end{acknowledgements}

%
%

\bibliographystyle{aa} 
\bibliography{Ref}
%
%
%
%
%
%
%
%

\begin{appendix}

\section{eFEDS-DOG Catalog Information}

In Table~\ref{table:1}, we provide a list of column names for our eFEDS-DOG catalogs. 
These catalogs encompass a range of photometric data, including ID, R.A., Dec., magnitudes in each band, optical spectral index, MIR and optical classification, observed flux between 0.5--2 keV, absorption-corrected intrinsic luminosity between 0.5--2 keV, redshift, redshift grade, absorption non-corrected luminosity between 0.5--2 keV, hydrogen column density, monochromatic luminosity at rest-frame 6 ${\rm \mu}$m, and photon index. 
Additionally, the catalog includes information on errors, upper and lower flags associated with each parameter.

The eFEDS-DOG samples were meticulously selected through the combination of catalogs utilizing HSC, VIKING, unWISE, and the eFEDS clean sample (refer to Section~\ref{DA:SS}). 
In total, the catalog comprises 65 eFEDS-DOGs, providing a comprehensive dataset for further analysis.

\begin{table*}
\caption{Column name, units and description of the our eFEDS-DOG catalogs.}\label{table:1}      
\centering          
\begin{tabular}{l c l }     
\hline\hline       
Name & Units & Description \\
\hline                    
   HSC\_ID  &        & HSC JHHMMSS.SS$\pm$DDMMSS.S \\  
   HSC\_RA & degree & R.A. from HSC coordintes    \\
   HSC\_DEC & degree & Dec. from HSC coordintes    \\
   HSC\_g & AB mag & HSC $g$-band magnitude \\
   HSC\_g\_err & AB mag & HSC $g$-band magnitude error \\
   HSC\_r & AB mag & HSC $r$-band magnitude \\
   HSC\_r\_err & AB mag & HSC $r$-band magnitude error \\
   HSC\_i & AB mag & HSC $i$-band magnitude \\
   HSC\_i\_err & AB mag & HSC $i$-band magnitude error \\
   HSC\_z & AB mag & HSC $z$-band magnitude \\
   HSC\_z\_err & AB mag & HSC $z$-band magnitude error \\
   HSC\_y & AB mag & HSC $y$-band magnitude \\
   HSC\_y\_err & AB mag & HSC $y$-band magnitude error \\
   VIKING\_Z & AB mag & VIKING $Z$-band magnitude \\
   VIKING\_Z\_err & AB mag & VIKING $Z$-band magnitude error\\
   VIKING\_Y & AB mag & VIKING $Y$-band magnitude \\
   VIKING\_Y\_err & AB mag & VIKING $Y$-band magnitude error\\
   VIKING\_J & AB mag & VIKING $J$-band magnitude \\
   VIKING\_J\_err & AB mag & VIKING $J$-band magnitude error\\
   VIKING\_H & AB mag & VIKING $H$-band magnitude \\
   VIKING\_H\_err & AB mag & VIKING $H$-band magnitude error\\
   VIKING\_Ks & AB mag & VIKING $Ks$-band magnitude \\
   VIKING\_Ks\_err & AB mag & VIKING $Ks$-band magnitude error\\
   WISE\_W1 & AB mag & WISE $W1$-band magnitude \\
   WISE\_W1\_err & AB mag & WISE $W1$-band magnitude error\\
   WISE\_W2 & AB mag & WISE $W2$-band magnitude \\
   WISE\_W2\_err & AB mag & WISE $W2$-band magnitude error\\
   WISE\_W2\_U\_F & & Upper limit flag WISE\_W2\\ 
   WISE\_W3 & AB mag & WISE $W3$-band magnitude \\
   WISE\_W3\_err & AB mag & WISE $W3$-band magnitude error\\
   WISE\_W3\_U\_F & & Upper limit flag WISE\_W3\\ 
   WISE\_W4 & AB mag & WISE $W4$-band magnitude \\
   WISE\_W4\_err & AB mag & WISE $W4$-band magnitude error\\
   MIR\_CLASS & & MIR classification (0: unclassified DOGs, 1: bump DOGs, 2: PL DOGs)\\
   ALPHA\_GRIZY & & $\alpha^{\rm opt}_{grizy}$\\
   ALPHA\_GRI & & $\alpha^{\rm opt}_{gri}$\\
   OPTICAL\_CLASS & & Optical classification (0: normal, 1: BluDOG-like)\\
   ID\_SRC & & ID of the sources in the eFEDS main X-ray catalog \citep{2022AandA...661A...1B}\\
   FOBS0520 & ${\rm erg\ s^{-1}\ cm^{-2}}$ & Observed flux in observed-frame 0.5-2.0 keV from \cite{2022AandA...661A...5L}\\
   FOBS0520\_err & ${\rm erg\ s^{-1}\ cm^{-2}}$ & Observed flux error in observed-frame 0.5-2.0 keV from \cite{2022AandA...661A...5L}\\
   Log10LINT0520\_ABSCORR & ${\rm erg\ s^{-1}}$ & $\lxsoftabs$ from \cite{2022AandA...661A...5L}\\
   Log10LINT0520\_ABSCORR\_err & ${\rm erg\ s^{-1}}$ & Error of $\lxsoftabs$ from \cite{2022AandA...661A...5L}\\
   REDSHIFT & & Redshift of the optical counterpart from \cite{2022AandA...661A...3S}\\
   REDSHIFT\_GRADE & & Redshift Grade from \cite{2022AandA...661A...3S}\\
   Log10L0520\_ABSNONCORR & ${\rm erg\ s^{-1}}$ & Absorption non-corrected luminosity in rest-frame 0.5-2.0 keV ($\lxsoftabsnoncorr$)\\
   Log10L0520\_ABSNONCORR\_err & ${\rm erg\ s^{-1}}$ & $\lxsoftabsnoncorr$ error\\
   Log10NH & ${\rm cm^{-2}}$ & Hydrogen column density ($N_{\rm H}$) from \cite{2022AandA...661A...5L}\\   
   Log10L6\_CIGALE & ${\rm erg\ s^{-1}}$ & monochromatic luminosity at rest-frame 6 ${\rm \mu m}$ ($L_{\rm 6\ \mu m}$) estimated by CIGALE\\
   Log10L6\_CIGALE\_err & ${\rm erg\ s^{-1}}$ & $L_{\rm 6\ \mu m}$ error\\
   Photon\_index & & Photon index ($\Gamma$) from \cite{2022AandA...661A...5L}\\ 
   Hard\_Detection & & Hard X-ray detection result (Y: ``yes'', N: ``no'')\\
\hline                  
\end{tabular}
\tablefoot{In lower and upper limit flag, 0 and 1 denote ``no'' and ``yes'', respectively. The prefix of ``Log10'' represents the values in log-scale.
}
\end{table*}

\section{Parameters for CIGALE Fitting}\label{sec:appendix_CIGALE}

To decompose the $L_{\rm 6\mu m}$ of the AGN from the total $L_{\rm 6\mu m}$, we employ the Code Investigating GAlaxy Emission \citep[CIGALE, version 2022.1;][]{2005MNRAS.360.1413B, 2009A&A...507.1793N, 2019A&A...622A.103B}. 
CIGALE performs spectral energy distribution (SED) fitting in a self-consistent framework, balancing UV/optical absorption and IR emission.

We adopt a delayed star formation history \citep[SFH;][]{2015AandA...576A..10C}, parameterized by the e-folding times of the main stellar population ($\tau_{\rm main}$) and late starburst population ($\tau_{\rm burst}$), the mass fraction of the late burst population ($f_{\rm burst}$), and the ages of the main stellar population (${\rm Age_{main}}$) and the late burst population (${\rm Age_{burst}}$). 
For the stellar population, we assume the initial mass function of \cite{2003PASP..115..763C}, solar metallicity, and a 10 Gyr separation between young and old stellar populations (${\rm Age_{separation}}$).

The nebular emission model from \cite{2011MNRAS.415.2920I} is characterized by the ionization parameter ($U$), the fraction of Lyman continuum photons escaping the galaxy ($f_{\rm esc}$), the fraction absorbed by dust ($f_{\rm dust}$), and the line width. 
Dust attenuation is modeled using the modified approach presented in \cite{2019A&A...622A.103B}. 
The attenuation for the continuum follows the model by \cite{2000ApJ...533..682C}, with extensions by \cite{2002ApJS..140..303L} for wavelengths between the Lyman break and 1500 \AA. 
Emission lines are attenuated using a Milky Way extinction law with $R_V = 3.1$ \citep[][]{1989ApJ...345..245C}, and we adopt $E(B-V)_{\mathrm {continuum}} = 0.44E(B-V)_{\mathrm{line}}$ as suggested by \cite{2000ApJ...533..682C}.

For the AGN emission, we utilize the SKIRTOR model, which incorporates geometric parameters of the AGN and accounts for extinction by polar dust. 
The parameters include the average edge-on optical depth at 9.7 $\mu$m ($\tau_{9.7}$), torus density parameters ($p$ and $q$; \cite{2016MNRAS.458.2288S}), the opening angle of the torus ($oa$), the ratio of the maximum to minimum radii of the torus ($R_{\rm ratio}$), the fraction of dust mass inside clumps ($M_{\rm cl}$), the inclination angle ($i$), the AGN fraction ($f_{\rm AGN}$), the extinction law, the color excess of the polar dust ($E(B-V)^{\rm AGN}_{\rm polar dust}$), the dust temperature ($T^{\rm AGN}_{\rm polar dust}$), and the emissivity index of the polar dust.

For BluDOG-like objects, the observed magnitudes in the $g$- and $r$-bands are treated as upper limits to account for the effects of broad emission lines with large equivalent widths \citep[][]{2022ApJ...941..195N}. In the SED fitting of these BluDOGs, we use the parameter set described in Table 8 of \cite{2022ApJ...941..195N}.

For eFEDS-DOGs without BluDOG-like objects, we utilize a parameter set shown in Table~\ref{table:CIGALE_parameter_set}.

\begin{table}
\caption{Parameters adopted in the CIGALE fit.}             
\label{table:CIGALE_parameter_set}      
\centering          
\begin{tabular}{c c}     
\hline\hline       
Parameter & Value \\
\hline                    
   \multicolumn{2}{c}{Delayed SFH \citep[][]{2015AandA...576A..10C}}\\\hline
   $\tau_{\rm main}$ [Myr]          & 100, 250, 500\\
   $\tau_{\rm burst}$ [Myr]         & 10, 50\\
   $f_{\rm burst}$                  & 0.0, 0.5, 0.99\\
   ${\rm Age}_{\rm main}$ [Myr]     & 500, 800, 1000\\
   ${\rm Age}_{\rm burst}$ [Myr]    & 1, 5, 10\\\hline
   \multicolumn{2}{c}{Single stellar population \citep[][]{2003MNRAS.344.1000B}}\\\hline
   IMF                              & \cite{2003PASP..115..763C}\\
   Metallicity                      & 0.02\\
   ${\rm Age}_{\rm separation}$ [Myr] & 10\\\hline
   \multicolumn{2}{c}{Nebular emission \citep[][]{2011MNRAS.415.2920I}}\\\hline
   $\log U$                         & -2.0\\
   $f_{\rm esc}$                    & 0.0\\
   $f_{\rm dust}$                   & 0.0\\
   Line width [km s$^{-1}$]         & 300.0\\\hline
   \multicolumn{2}{c}{Dust attenuation \citep[][]{2000ApJ...533..682C}}\\\hline
   $E(B-V)_{\rm line}$              & 0.03, 0.1, 0.3, 1, 3\\
   $f_{E(B-V)}$                     & 0.44\\
   $\lambda_{\rm UV,bump}$ [nm]     & 217.5\\
   ${\rm FWHM}_{\rm UV,bump}$ [nm]  & 35.0\\
   $A_{\rm UV,bump}$ [nm]           & 0.0\\
   $\delta$                         & 0.0\\
   Extinction law of emission lines & the Milky Way\\
   $R_V$                            & 3.1\\\hline
   \multicolumn{2}{c}{Dust emission \citep[][]{2014ApJ...784...83D}}\\\hline
   AGN fraction                     & 0.0\\
   $\alpha_{\rm IR,AGN}$            & 0.25, 1.0, 4.0\\\hline
   \multicolumn{2}{c}{AGN model \citep[][]{2016MNRAS.458.2288S}}\\\hline
   $\tau_{9.7}$                     & 3, 7, 11\\
   $p$                              & 1.0\\
   $q$                              & 1.0\\
   $oa$ [deg]                       & 10, 30, 50, 70, 80\\
   $R_{\rm ratio}$                  & 20\\
   $M_{\rm cl}$                     & 0.97\\
   $i$ [deg]                        & 0, 20, 40, 60, 80, 90\\
   $f_{\rm AGN}$                    & 0.1,0.5,0.99\\
   Extinction law of polar dust     & \cite{2000ApJ...533..682C}\\
   $E(B-V)^{\rm AGN}_{\rm polar\ dust}$  & 0.03, 0.1, 0.3, 1, 3\\
   $T^{\rm AGN}_{\rm polar\ dust}$ [K]  & 30, 100, 300, 1000\\
   Emissivity of polar dust         & 1.6\\
\hline                  
\end{tabular}
\end{table}

\section{Criteria of upper-limit of $\lsix$}\label{sec:appendix_L6}
To estimate $L_{6\mathrm{\mu m}}$ for each detection case and redshift, we classified eFEDS-DOGs according to the categories outlined in Table~\ref{tab:L6up}. 
Additionally, we summarize the criteria for the upper limit of $L_{\mathrm{6 \mu m}}$.

(1) For objects with detections in the $W2$, $W3$, and $W4$-bands, we fit their MIR SEDs using a power-law function. 
Out of the 65 eFEDS-DOGs, 44 sources are detected in all three MIR bands. 
We estimate $L_{\mathrm{6 \mu m}}$ from the best-fit power-law function, and no upper flag is assigned to $L_{\mathrm{6 \mu m}}$.

(2) One eFEDS-DOG shows non-detection only in the $W2$-band. For this source, we extrapolate the flux density at rest-frame 6 $\mu$m ($f_{\mathrm{6 \mu m}}$) using a power-law extrapolation from the slope between W3 and W4. Since the extrapolated flux density at W2 ($f_{W2}^{\mathrm{Estimated}}$) is fainter than the limiting flux density at W2 ($f_{W2}^{\mathrm{Estimated}} < f_{W2}^{\mathrm{limit}}$), no upper flag is assigned to $L_{\mathrm{6 \mu m}}$.

(3) Out of 65 sources, 18 eFEDS-DOGs were undetected in the $W3$-band. 
Among them, three show that their estimated $W3$-band fluxes ($f^{\mathrm{Estimated}}_{W3}$), interpolated by $W2$ and $W4$-bands fluxes, are fainter than the $W3$-band limiting flux ($f_{W3}^{\mathrm{limit}}$). 
In this case, we estimate $f_{\mathrm{6 \mu m}}$ using a power-law interpolation 
between W2 and W4, and no upper flag is assigned to $L_{\mathrm{6 \mu m}}$.

(4) For the remaining 15 eFEDS-DOGs undetected in the $W3$-band,
the simple interpolation between W2 and W4 results in $f^{\mathrm{Estimated}}_{W3}>f_{W3}^{\mathrm{limit}}$. 
In this case, we use $f_{W3}^{\mathrm{limit}}$ to estimate $f_{\mathrm{6 \mu m}}$, employing a power-law interpolation between $f_{W2}$--$f_{W3}^{\mathrm{limit}}$ or $f_{W3}^{\mathrm{limit}}$--$f_{W4}$. 
An upper flag is assigned to these $L_{\mathrm{6 \mu m}}$.

(5) Two eFEDS-DOGs are undetected both in the $W2$ and $W3$-bands. We utilize the limiting flux density of $W2$ and $W3$-bands, and $f_{\mathrm{6 \mu m}}$ is estimated using a power-law interpolation between $f_{W2}^{\mathrm{limit}}$--$f_{W3}^{\mathrm{limit}}$ or $f_{W3}^{\mathrm{limit}}$--$f_{W4}$. An upper flag is assigned to these $L_{\mathrm{6 \mu m}}$.

\begin{table*}
\caption{Upper Limits of $L_{6\ \rm{\mu m}}$.}             
\label{tab:L6up}      
\centering          
\begin{tabular}{c l l l r l } 
\hline\hline    
& Condition & Criteria  & & Upper Limit Flag    & \# eFEDS-DOGs\\
\hline
(1) & \multicolumn{3}{l}{3-band detected} & No & 44\\ 
(2) & Only $W2$-band undetected & \& $(1+z)\times\lambda_{\rm 6\mu m} < \lambda_{W3}$ & \& $f^{\rm Estimated}_{W2} < f^{\rm limit}_{W2}$ & No & 1\\
(3) & Only $W3$-band undetected & \& $f^{\rm Estimated}_{W3} < f^{\rm limit}_{W3}$ & & No & 3\\
(4) & Only $W3$-band undetected & \& $f^{\rm limit}_{W3} < f^{\rm Estimated}_{W3}$ & & Yes & 15\\
(5) & \multicolumn{3}{l}{$W2$ \& $W3$-bands undetected} & Yes & 2\\
\hline
Total&&&&&65\\
\hline
\end{tabular}
\tablefoot{Condition: Statuses of three MIR bands, Criteria: Definitions and explanations for the terms used in the table, Upper Limit flag: We attached/did not attach upper limit flags if objects have these Condition, \# eFEDS-DOGs: Number of eFEDS-DOGs classified under these Condition.
$\lambda_{W3}$, $\lambda_{\rm 6\ \mu m}$, $f^{\rm Estimated}_{W2}$, $f^{\rm limit}_{W2}$, $f^{\rm Estimated}_{W3}$, and $f^{\rm limit}_{W3}$ denote $W3$-band wavelength, 6 $\mu$m, estimated flux at $W2$-band, $W2$-band limiting flux, estimated flux at $W3$-band, and $W3$-band limiting flux, respectively. Each estimated flux is calculated as follows:\\
(1): Best-fit function from MIR 3-band fitting (see Section~\ref{MIR_CLASS_DOGs}).\\
(2): Interpolated function from $W3$- and $W4$-bands data.\\
(3): Interpolated or extrapolated function from $W2$- and $W4$-bands data.\\
(4): Interpolated function from $W2$-, $W3$-, and $W4$-bands data.\\
(5): Interpolated function from $W2$-, $W3$-, and $W4$-bands data.
}
\end{table*}

\end{appendix}

\end{document}